\documentclass[twocolumn]{aastex631}
\shorttitle{Metallicity-PAH relation of MIR galaxies}
\shortauthors{Shim et al.}
\graphicspath{{./}{figures/}}

\begin{document}

\title{Metallicity-PAH Relation of MIR-selected Star-forming Galaxies in AKARI North Ecliptic Pole-wide Survey}

\correspondingauthor{Hyunjin Shim}
\email{hjshim@knu.ac.kr}

\author[0000-0002-4179-2628]{Hyunjin Shim}
\affiliation{Department of Earth Science Education, Kyungpook National University, 80 Daehak-ro, Buk-gu, Daegu 41566, Republic of Korea} 

\author[0000-0003-3428-7612]{Ho Seong Hwang}
\affiliation{Astronomy Program, Department of Physics and Astronomy, Seoul National University, 1 Gwanak-ro, Gwanak-gu, Seoul 08826, Republic of Korea}
\affiliation{SNU Astronomy Research Center, Seoul National University, 1 Gwanak-ro, Gwanak-gu, Seoul 08826, Republic of Korea}

\author[0000-0002-2770-808X]{Woong-Seob Jeong}
\affiliation{Korea Astronomy and Space Science Institute, 776 Daedeok-daero, Yuseong-gu, Daejeon 34055, Republic of Korea}

\author[0000-0002-3531-7863]{Yoshiki Toba}
\affiliation{National Astronomical Observatory of Japan, 2-21-1 Osawa, Mitaka, Tokyo 181-8588, Japan}
\affiliation{Academia Sinica Institute of Astronomy and Astrophysics, 11F of Astronomy-Mathematics Building, AS/NTU, No.1, Section 4, Roosevelt Road, Taipei 10617, Taiwan} 
\affiliation{Research Center for Space and Cosmic Evolution, Ehime University, 2-5 Bunkyo-cho, Matsuyama, Ehime 790-8577, Japan}

\author[0000-0002-3560-0781]{Minjin Kim}
\affiliation{Department of Astronomy and Atmospheric Sciences, Kyungpook National University, 80 Daehak-ro, Buk-gu, Daegu 41566, Republic of Korea}

\author[0000-0002-6925-4821]{Dohyeong Kim}
\affiliation{Department of Earth Sciences, Pusan National University, 2, Busandaehak-ro 63 beon-gil, Geumjeong-gu, Busan 46241, Republic of Korea}

\author[0000-0002-4362-4070]{Hyunmi Song}
\affiliation{Department of Astronomy and Space Science, Chungnam National University, 99 Daehak-ro, Yuseong-gu, Daejeon 34134, Republic of Korea}

\author[0000-0001-7228-1428]{Tetsuya Hashimoto}
\affiliation{Department of Physics, National Chung Hsing University, 145 Xingda Rd., South Dist., Taichung City, 402, Taiwan}

\author[0000-0002-6660-9375]{Takago Nakagawa}
\affiliation{Institute of Space and Astronautical Science, Japan Aerospace Exploration Agency, 3-1-1 Yoshinodai, Chuo-ku, Sagamihara, Kanagawa 252-5210, Japan}

\author[0000-0001-6652-1069]{Ambra Nanni}
\affiliation{National Centre for Nuclear Research, ul. Pasteura 7, 02-093 Warszawa, Poland}

\author[0000-0002-7300-2213]{William J. Pearson}
\affiliation{National Centre for Nuclear Research, ul. Pasteura 7, 02-093 Warszawa, Poland}

\author{Toshinobu Takagi}
\affiliation{Japan Science Forum, 3-2-1, Kandasurugadai, Chiyoda-ku, Tokyo 101-0062, Japan}



\begin{abstract}

	We investigate the variation 
	in the mid-infrared spectral energy distributions
	of 373 low-redshift ($z<0.4$) star-forming galaxies, 
	which reflects 
	a variety of polycyclic aromatic hydrocarbon (PAH) 
	emission features.
	The relative strength of PAH emission 
	is parameterized as $q_\mathrm{PAH}$, 
	which is defined as the mass fraction of PAH particles
	in the total dust mass. 
	With the aid of continuous mid-infrared photometric data points 
	covering 7-24\,$\mu$m and far-infrared flux densities, 
	$q_\mathrm{PAH}$ values are derived 
	through spectral energy distribution fitting. 
	The correlation between $q_\mathrm{PAH}$ and
	other physical properties of galaxies, 
	i.e., gas-phase metallicity ($12+\mathrm{log(O/H)}$),
	stellar mass, and specific star-formation rate (sSFR) 
	are explored. 
	As in previous studies, $q_\mathrm{PAH}$ values
	of galaxies with high metallicity 
	are found to be higher than those with low metallicity.
	The strength of PAH emission is also positively
	correlated with the stellar mass 
	and negatively correlated with the sSFR.
	The correlation between $q_\mathrm{PAH}$
	and each parameter still exists
	even after the other two parameters are fixed.
	In addition to the PAH strength, 
	the application of metallicity-dependent gas-to-dust mass ratio
	appears to work well 
	to estimate gas mass 
	that matches the observed relationship 
	between molecular gas and physical parameters. 
	The result obtained will be used to calibrate the observed 
	PAH luminosity-total infrared luminosity relation,
	based on the variation of MIR-FIR SED, 
	which is used in the estimation of hidden star formation. 
\end{abstract}



\section{Introduction} \label{sec:intro}

Mid-infrared spectra of star-forming galaxies are 
frequently dominated by strong emission features 
at 3.3, 6.2, 7.7, 8.6, 11.3, and 12.7\,$\mu$m
from polycyclic aromatic hydrocarbon (PAH) molecules. 
PAHs contain more than tens to thousands carbon atoms 
as well as peripheral hydrogen atoms that produce
emission through bending and stretching
\citep[see][for a review]{2008ARA&A..46..289T}, 
and are considered to be an important component 
of the interstellar dust population. 
Since dust plays a significant role in the star-formation process 
by enhancing the formation of molecular hydrogen 
and being an efficient coolant of interstellar medium (ISM), 
PAHs are closely related to the star-formation activity. 
PAH emission features are observed    
in a broad range of astrophysical scales, 
including Galactic and extragalactic star-forming regions 
\citep[e.g.,][]{2018MNRAS.481.5370M, 2022MNRAS.509.3523K},
therefore the possibility of using PAH luminosity 
as star-formation rate (SFR) indicator
has been intensively explored 
\citep[e.g.,][]{2004ApJ...613..986P, 2007ApJ...666..870C, 2012ApJ...756...95L, 
2016ApJ...818...60S, 2019ApJ...884..136X, 2020ApJ...905...55L}. 
The ubiquitous existence of PAH emission features
in high-redshift (up to $z\sim2$) galaxies 
\citep[e.g.,][]{2005A&A...434L...1E, 2005ApJ...625L..83L, 2005ApJ...628..604Y, 2020NatAs...4..339L}
demonstrates potential for 
rest-frame MIR photometric or spectroscopic observations, 
such as probing the cosmic star-formation history 
in early universe \citep{2004ApJS..154..112L, 2010A&A...514A...6G}  
and investigating the contribution of star formation 
in active galactic nuclei (AGN) host systems \citep{2020A&A...639A..43A}. 

In order to establish the PAH-SFR correlation, 
the variation in PAH emissions
in different ISM environments needs to be understood beforehand. 
One of the most well-studied factors 
relevant to the PAH variation is metallicity. 
It has been reported that PAH emission features are 
less prominent 
in low-metallicity environments 
\citep[e.g.,][]{2005ApJ...628L..29E, 2006A&A...446..877M, 2006ApJ...639..157W, 2008ApJ...679..310G}.
The emission features show negative correlation  
with the hardness of the radiation field 
(i.e., [Ne\,\textsc{iii}]/[Ne\,\textsc{ii}] ratios). 
In some cases, 
MIR spectra of ultraluminous infrared galaxies
lack PAH emission features \citep{2006ApJ...641..133D}, 
indicating that these systems are dominated by AGN 
or are in a vigorous merging process
\citep[][]{2013PASJ...65..103Y, 2017MNRAS.472...39M}.
It is suggested that 
PAH destruction may be more effective in low metallicity systems
than in high metallicity systems, 
due to a hard radiation field 
and inefficient gas cooling. 
Another possibility is a low production of PAHs 
in low metallicity systems with low carbon abundance 
\citep{2005AIPC..761..103D},
which is related to the evolution of carbon-rich TP-AGB stars 
\citep{2008ApJ...672..214G}.
The trend of PAH strength being dependent on the metallicity 
is also observed in galaxies at $z\sim2$ 
\citep{2017ApJ...837..157S}, 
affecting a scaling relation between MIR and total IR luminosity 
that is frequently employed to estimate 
an attenuation-free SFR density 
based on the flux density measured using a single MIR photometric band. 

Observations of local galaxies that are clearly detected 
in the MIR and FIR bands
are used as the basis for building a dust emission model 
at MIR-FIR wavelengths.
Based on the
Spitzer and submillimeter observations
of the SIRTF Nearby Galaxies Survey 
\citep[SINGS;][]{2003PASP..115..928K},  
\citet{2007ApJ...663..866D}
have suggested that the dust-to-gas ratio in star-forming galaxies
is dependent on the metallicity,
and the PAH strength correlates with the metallicity as well.
To parameterize the relative strength of PAH emission 
with respect to the total IR luminosity, 
\citet{2007ApJ...657..810D} dust emission model 
have defined the PAH index $q_\mathrm{PAH}$,
i.e., 
fraction of dust mass 
locked within PAH molecules (with less than $10^3$ C atoms).
In case of SINGS galaxies, 
the median $q_\mathrm{PAH}$ value of low-metallicity galaxies
($12+\mathrm{log(O/H)}<8.1$)
is 1.0\,\%, while the value for
high-metallicity galaxies ($12+\mathrm{log(O/H)}\ge8.1$)
is 3.55\,\%. 
The parameter $q_\mathrm{PAH}$ 
has also been applied to investigate 
MIR-FIR spectral energy distribution (SED) of 
Herschel Reference Survey galaxies
\citep{2014A&A...565A.128C}. 
By presenting a positive correlation 
between $q_\mathrm{PAH}$ and oxygen abundance 
in a range of $8.2 < 12+\mathrm{log(O/H)} < 8.5$, 
\citet{2014A&A...565A.128C} have suggested that 
$q_\mathrm{PAH}$ factor may provide metallicity constraint 
to galaxies ranging one-third solar metallicity 
to solar metallicity.

The aforementioned works 
are based on the studies of 
local ($<30$\,Mpc) galaxies
with a limited number of MIR photometry points.
However, since the PAH spectrum is affected by several factors 
such as particle size and charge distribution
\citep{2007ApJ...656..770S, 2020MNRAS.494..642M},
flux densities in multiple MIR bands are required 
to better constrain the PAH luminosity 
and to identify PAH-luminous galaxies \citep[e.g.,][]{2010A&A...514A...5T}.
For example, in estimating $q_\mathrm{PAH}$ using
SED fitting \citep{2019PASJ...71...27K}, 
it is indicated that the addition of the 
11, 15, and 18\,$\mu$m bands of the AKARI/IRC
is more advantageous than the use of single 22 or 24\,$\mu$m band. 
Unfortunately, the majority of the galaxies 
described in \citet{2019PASJ...71...27K} 
lack spectroscopic redshifts, 
and the use of photometric redshifts in SED fitting 
increases uncertainties in the derived physical parameters.
MIR multiband survey data, complemented with 
optical spectroscopic observations
and ancillary multiwavelength data 
especially equipped with FIR coverage, 
would allow us to probe the PAH strength variation of 
MIR-selected galaxies 
in terms of galaxy physical properties such as metallicity, 
SFR, and stellar mass. 

In this paper, 
we present an analysis on the correlation between PAH abundance 
and gas-phase metallicity (i.e., oxygen abundance) 
of MIR-selected galaxies at $z<0.4$, 
using the continuous MIR photometry at 7-24\,$\mu$m 
in addition to the FIR photometry at $\ge250$\,$\mu$m. 
Section~\ref{sec:data} describes the characteristics of data we use 
and sample selection criteria.
How metallicity and PAH mass fraction,
as well as other physical properties of galaxies are measured 
is described in Sections~\ref{sec:metal} and \ref{sec:sedfit},
respectively.
The results on the correlation between PAH 
and other factors are discussed in Section~\ref{sec:result}. 
Our work is summarized in Section~\ref{sec:summary}.
Throughout the paper, we use 
Planck 2018 cosmological parameters
for a flat $\Lambda$CDM model
\citep[][$\Omega_{\rm m, 0}=0.315$, $H_0=67.4\,$km\,s$^{-1}$\,Mpc$^{-1}$]{2020A&A...641A...6P}.

\section{Data and Sample}   \label{sec:data}

\subsection{NEP-wide Multiwavelength Survey}

The $\sim5.4$\,deg$^2$ region around the North Ecliptic Pole (NEP)
has been observed by the infrared space telescope
AKARI \citep{2007PASJ...59S.369M},
using nine near- to mid-IR filters of the Infrared Camera
(IRC; N2, N3, N4, S7, S9W, S11, L15, L18W, L24)
providing continuous wavelength coverage
between 2 and 24\,$\mu$m
\citep[NEP-wide survey;][]{2012A&A...548A..29K}.
Numbers in the filter names represent
the central wavelengths of the filters,
while the ``W'' indicates wider filter width.
The AKARI MIR bands provide information
at wavelengths covering
strong PAH emission lines (7.7, 8.6, 11.3 and 12.7\,$\mu$m),
making them good tools to estimate the PAH luminosity.
\citet{2012A&A...548A..29K} and \citet{2021MNRAS.500.4078K}
have provided catalogs of the sources
detected in the AKARI/IRC.
The numbers of the detected sources are different
for different detection bands,
e.g., $\sim100,000$ in the NIR band (N3),
$\sim18,000$ in the S9W band and $\sim4000$ in the L24 band.
After excluding suspicious false detections
that are related to cosmic-ray hits and data reduction artifacts,
\citet{2021MNRAS.500.4078K}
have finalized the NEP-wide catalog
with 130,150 infrared sources that have detection in
at least one of the AKARI/IRC nine bands.
The $5\sigma$ flux limits
in N2, N3, N4, S7, S9W, S11, L15, L18W, and L24
are 15.4, 13.3, 13.6, 58.6, 67.3, 93.8,
133.1, 120.2, and 274.4\,$\mu$Jy, respectively.

Multiwavelength ancillary data sets are available in this region
\citep[see \citealt{2021MNRAS.500.4078K} for a summary;][]{2007ApJS..172..583H,
2014ApJS..214...20J, 2017PKAS...32..219P,
2018ApJS..234...38N, 2020MNRAS.498..609H,
2020MNRAS.498.5065S, 2021MNRAS.500.5024O}.
The available dedicated multiwavelength observations
and the legacy survey observations
\citep[using GALEX and WISE;][]{2017ApJS..230...24B, 2012yCat.2311....0C}
from ultraviolet to FIR wavelengths
allow the exploration of the physical properties of
MIR-selected galaxies,
including redshift, SFR, stellar mass,
and total IR luminosity.
For example, \citet{2021MNRAS.500.4078K}
have presented that 111,535 out of 130,150 AKARI sources
have counterparts in the deep optical images obtained by Subaru/HSC
\citep{2021MNRAS.500.5024O}.
Among them, 19,674 sources are flagged to be associated with
bad pixels, saturated pixels, and edges of the masked regions,
which results the number of reliable AKARI sources
with optical identification to be 91,861.
Photometric redshifts of these 91,861 sources
have been estimated by \citet{2021MNRAS.502..140H}
based on the template fitting using optical-NIR photometry.
Target selection for optical spectroscopic observations
in the following Section~\ref{sec:specobs}
utilizes the AKARI source catalog
matched with optical source catalog.

\begin{figure*}
\epsscale{1.1}
\plottwo{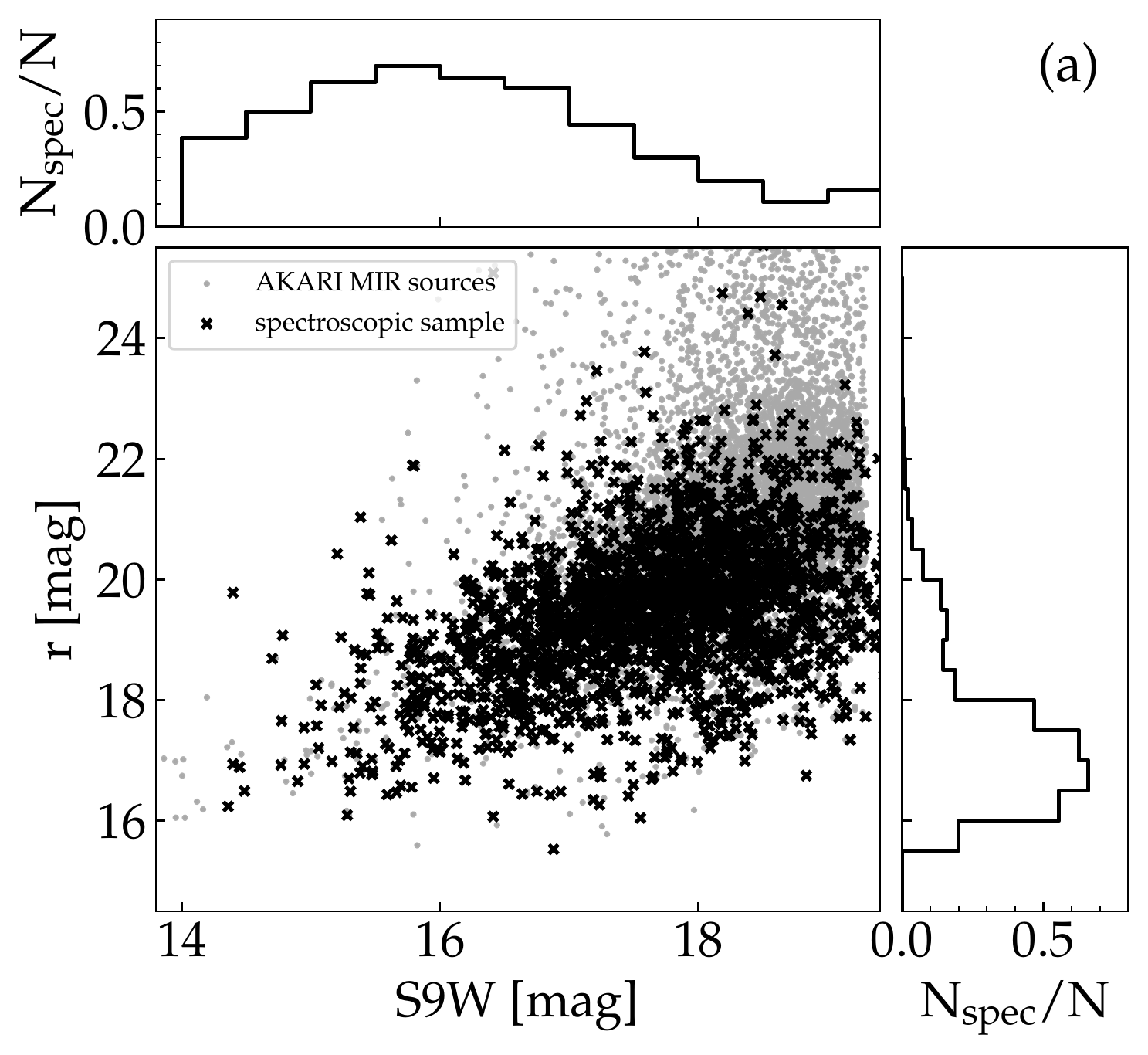}{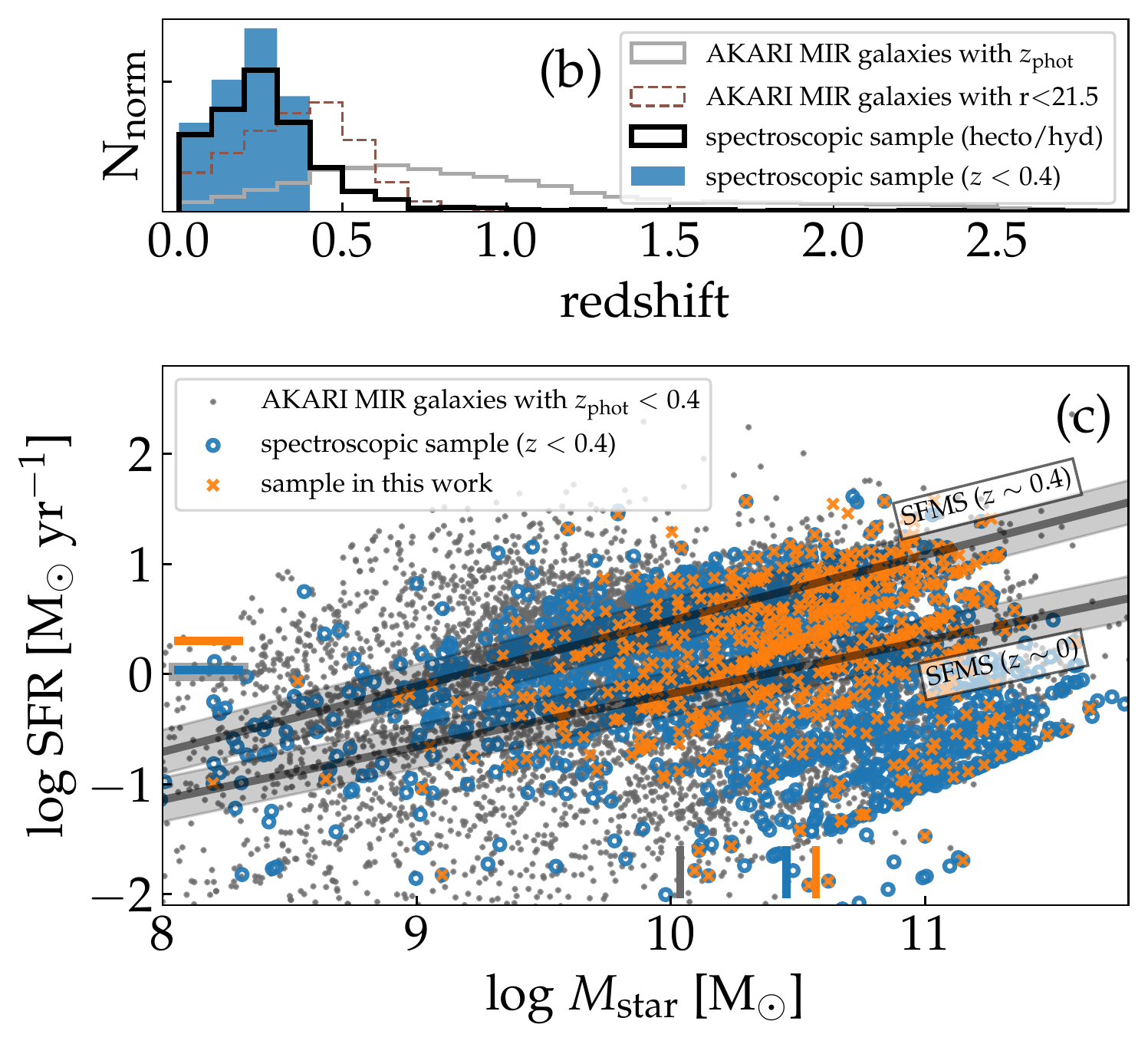}
\caption{\label{fig:specsample}
        (a) Comparison between the parent 
	AKARI MIR-selected sources
	(dots) and sources with optical spectroscopic observations
	(crosses). Histograms on the top and the right show
	the number fractions of the sources with optical spectra
	among the entire MIR-selected sources,
	i.e., the completeness of the spectroscopic observations,
	for different magnitude bins.
	Optical spectra are available for more than 50\% of the
	MIR-selected sources with $f_\mathrm{S9W}>0.4$\,mJy
	(corresponds to $f_\mathrm{S11}>0.6$\,mJy and/or
	$f_\mathrm{L15}>0.9$\,mJy).
	(b) Redshift distributions of
	(i) AKARI MIR galaxies with optical identification
	\citep[photometric redshifts from][]{2021MNRAS.502..140H},
	(ii) AKARI galaxies with $r$-band magnitude cut ($r<21.5$),
	(iii) MIR-selected sources targeted for optical spectroscopic observations
	using MMT/Hectospec and WIYN/Hydra,
	and (iv) $z_\mathrm{spec}<0.4$ sources studied in this work.
	All histograms are normalized to have equal areas.
	(c) Star-formation rate vs. stellar mass from SED fitting
	for AKARI galaxies with $z_\mathrm{phot}<0.4$ (dots), 
	spectroscopic sample of galaxies with $z_\mathrm{spec}<0.4$ 
	(open circles), 
	and subsample of spectroscopic sample 
	that also satisfy MIR and FIR detection criteria (crosses; 
	see Section~\ref{sec:sample} for sample selection in this work).
	Overplotted solid lines represent 
	star-forming main sequences at $z\sim0$ and $z\sim0.4$
	\citep{2014ApJS..214...15S} with $\pm0.2$\,dex scatter as shaded regions.
	Short vertical and horizontal bars with different colors 
	indicate median values of $\mathrm{log}\,M_\mathrm{star}$ 
	and $\mathrm{log}\,\mathrm{SFR}$ for different samples, respectively.
}
\end{figure*}

\subsection{Optical Spectroscopic Observations}
\label{sec:specobs}

Optical spectra of MIR-selected galaxies 
over the NEP-wide survey area 
have been 
obtained using the MMT/Hectospec 
and WIYN/Hydra multi-fiber spectrographs 
between 2008 and 2021, 
through the spectroscopic follow-up campaign of MIR sources.
The earlier data using MMT/Hectospec and WIYN/Hydra 
(observed by 2008) 
have been published by \citet{2013ApJS..207...37S}, 
where the targets are selected in 
11\,$\mu$m and 15\,$\mu$m band
(more than $10\sigma$ detection in both bands).
Later in 2020 and 2021, spectroscopic observations 
of 9\,$\mu$m-selected sources 
(with S9W magnitude brighter than 19.5\,mag)
have been carried out using MMT/Hectospec 
with the aim of increasing the number of MIR-selected sources
with spectroscopic redshifts,
which is essential for understanding the dependence of 
MIR-based star-formation activity as a function of galaxy environment
(Hwang, H. S. et al. 2022, in preparation).
The configurations for MMT/Hectospec observations 
in 2008 and 2021 are identical, 
i.e., using 270 lines mm$^{-1}$ grating to cover 3700-8500\,\AA\,
at a spectral resolution of $6\,$\AA\, 
with dispersion of 1.2\,\AA\,pixel$^{-1}$. 
The spectral resolution and dispersion of WIYN/Hydra spectra 
are similar to that of MMT/Hectospec. 
However, the quality of the WIYN/Hydra spectra 
in the wavelengths blueward of 4500\,\AA\,
is poor due to the low instrumental response.

Figure~\ref{fig:specsample}a shows
magnitude distribution of 
the parent AKARI MIR sources 
(91,861 sources from \citealt{2021MNRAS.500.4078K}) 
and the targets for optical spectroscopy 
in the $r$ vs. S9W plane.
The approximate magnitude limit 
in optical ($r$-band) wavelength is $r=21.5$\,mag,
although some sources fainter 
than the $r$-band magnitude limit
had a chance to occupy the spectroscopic fibers. 
The number fraction of the sources that have been selected 
for spectroscopic follow-up decreases as the $r$-band magnitude
increases. 
At magnitudes of $\mathrm{S9W}<17.4$ 
(i.e., $f_\mathrm{S9W}>0.4$\,mJy), 
about 50\% of the AKARI sources are 
followed up for optical spectroscopic observations. 
In S11 and L15 bands, 
the flux limits where the 50\% spectroscopic completeness
is reached 
are $f_\mathrm{S11}>0.6$\,mJy and $f_\mathrm{L15}>0.9$\,mJy,
respectively.
The redshift distributions of the parent photometric sample 
and the spectroscopic sample are compared 
in Figure~\ref{fig:specsample}b.
The parent sample plotted here represent 67,616 objects 
of which photometric redshifts are calculated 
and are classified as galaxies with better fits to galaxy template
than stellar template \citep{2021MNRAS.502..140H}. 
For the entire MIR sources,
the median value of the redshift distribution is 
$\langle z_\mathrm{phot}\rangle=0.82$. 
If the $r$-band magnitude cut is applied, 
the value decreases with the decrease of the magnitude limit:
$\langle z_\mathrm{phot}\rangle=0.38$,
0.27, and 0.18 for $r<21.5$, 20.5, and 19.5\,mag.  
Considering that the spectroscopic sample completeness
depends on the $r$-band magnitude, 
it is reasonable that the median redshift 
of the spectroscopic sample is lower than 
that of the photometric sample, 
i.e., $\langle z_\mathrm{spec} \rangle=0.22$.

The two representative physical properties, 
SFR and stellar mass, 
of the photometric and spectroscopic sample
(both limited to redshift $z<0.4$)
are illustrated in Figure~\ref{fig:specsample}c. 
Most sources are located on the previously known 
sequence of star-forming galaxies \citep[e.g.,][]{2014ApJS..214...15S}
at $0<z<0.4$. 
The median values of $\mathrm{log}\,M_\mathrm{star}$ 
and $\mathrm{log}\,\mathrm{SFR}$ for different samples 
show that spectroscopic sample is composed of massive galaxies, 
as the sample consists of $r$-band bright sources. 
The median stellar mass of the $z<0.4$ photometric sample galaxies
is $\langle\mathrm{log}\,M_\mathrm{star}[\mathrm{M}_\odot]\rangle=10$
while that of the spectroscopic sample galaxies is
$\langle\mathrm{log}\,M_\mathrm{star}[\mathrm{M}_\odot]\rangle=10.4$.
By applying Kolmogorov-Smirnov test, 
the stellar masses of the two samples 
are statistically different with a $p$-value less than 0.05. 
In case of the SFR, 
little difference is found for photometric and spectroscopic sample 
with the median values of 
$\langle\mathrm{log}\,\mathrm{SFR}[\mathrm{M}_\odot\,\mathrm{yr}^{-1}]\rangle=0.02$
and 0.04, respectively.

\subsection{Emission Line Flux Measurement}

\begin{figure*}
\epsscale{1.1}
\plottwo{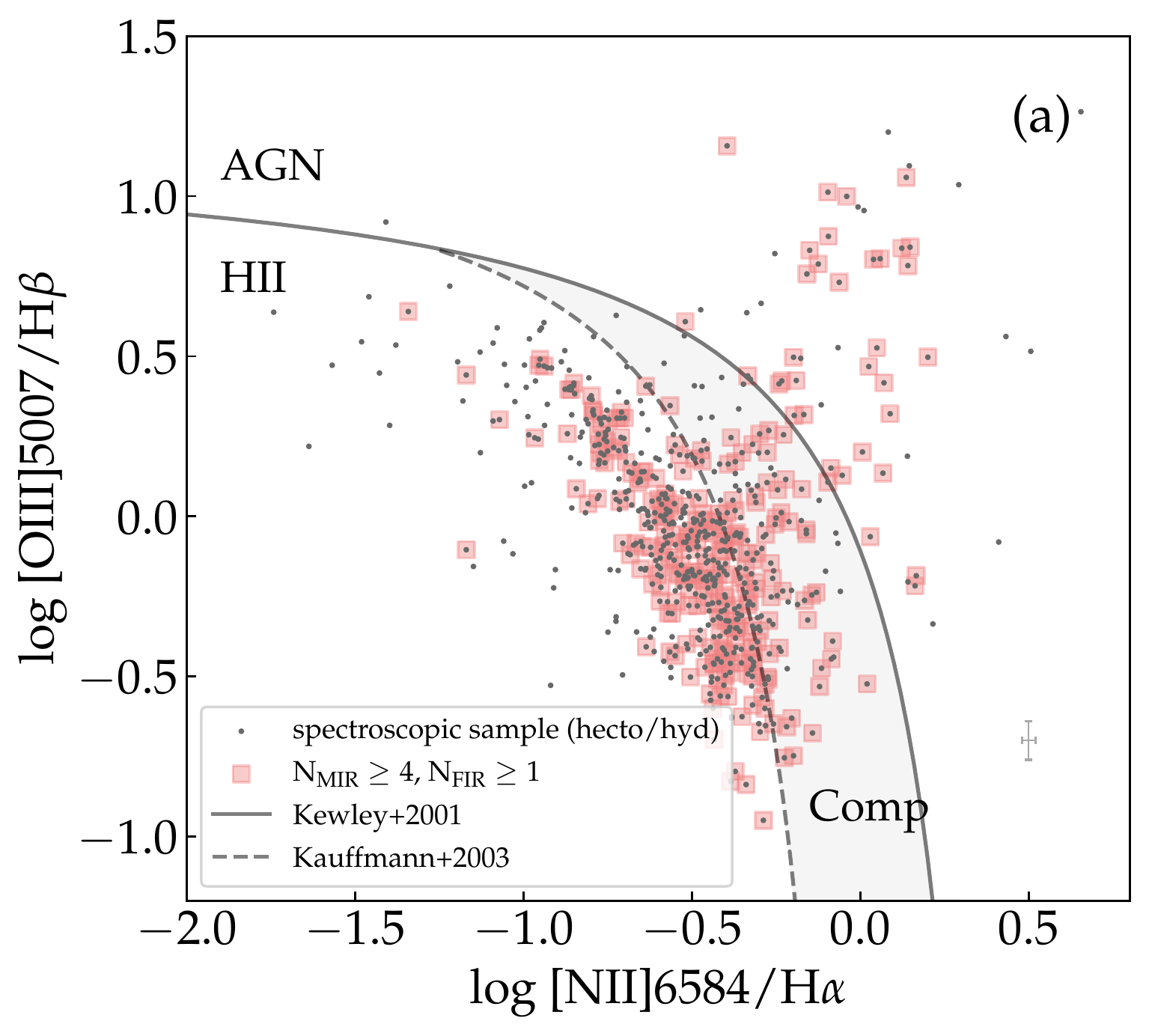}{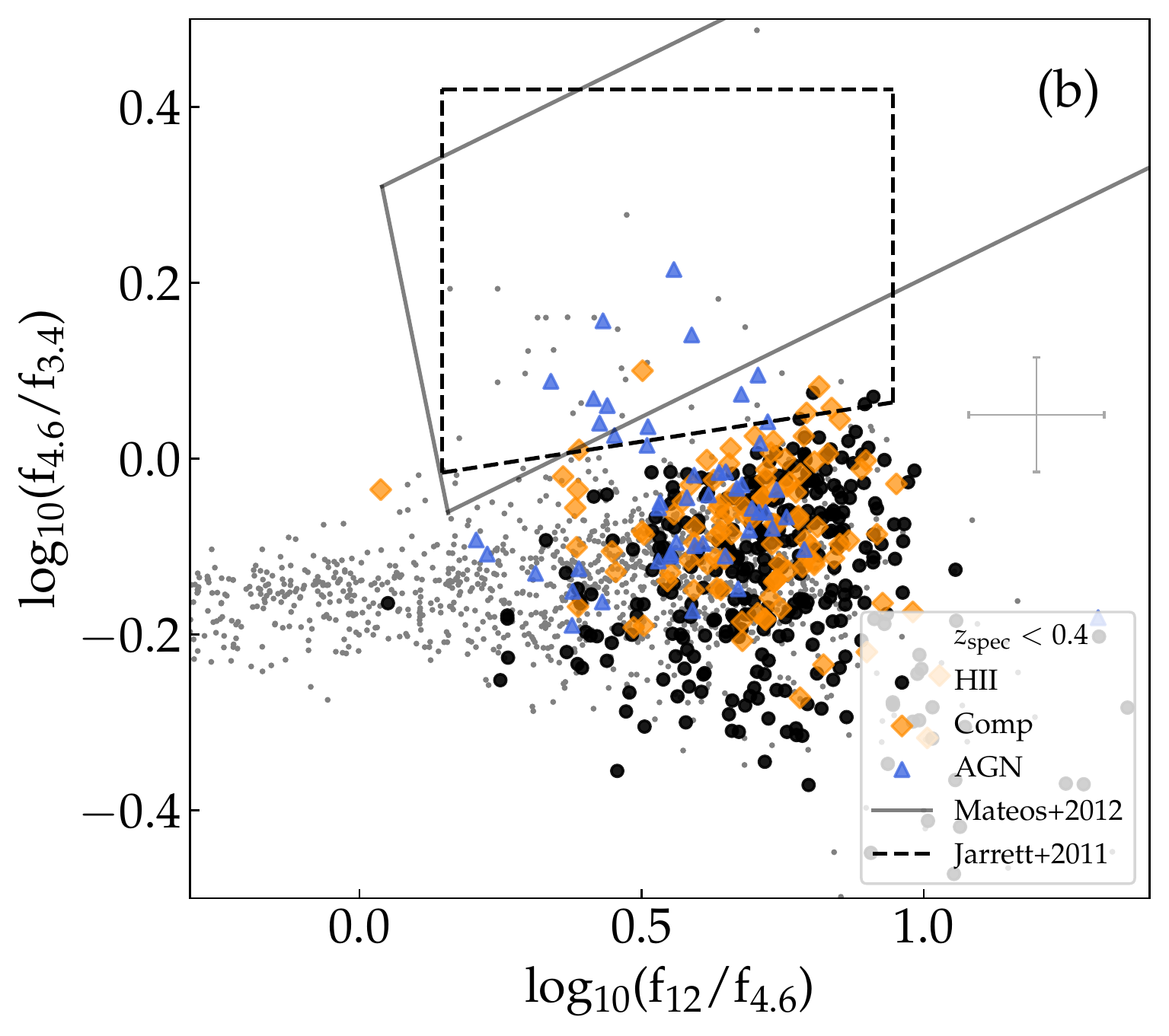}
\caption{\label{fig:bpt}
(a) BPT diagram \citep{1981PASP...93....5B}
for MIR-selected sources.
Objects with sufficient MIR and FIR photometric data points
(at least four in MIR wavelengths and one in FIR wavelengths)
are marked as coral squares.
Typical values of errors in $x$- and $y$-axes
are plotted as an error bar at the bottom right.
Overplotted are criteria to separate AGN and star-forming galaxies
(\citealt{2001ApJ...556..121K} and \citealt{2003MNRAS.346.1055K}
for solid and dashed lines, respectively).
(b) MIR color-color diagram for AGN selection
using WISE 3-bands (3.4, 4.6, and 12\,$\mu$m) photometry.
Dots are WISE-detected sources
with spectroscopic redshifts of $z<0.4$.
Objects that are classified by the BPT diagram
as star-forming (H\,\textsc{ii}), composite,
and AGN subgroups are plotted using different symbols.
Typical error bar in both axes is shown in the right.
Solid and dashed lines represent
AGN selection criteria
suggested by \citet{2012MNRAS.426.3271M}
and \citet{2011ApJ...735..112J}, respectively,
based on the
assumption that AGN contribution is described by
a power-law-shaped hot dust emission at 3-12\,$\mu$m.
}
\end{figure*}

The extracted 1D spectra are flux-calibrated 
using the spectra of spectrophotometric standard stars 
that are simultaneously observed by filling the fibers in each 
observation configuration. 
Once the spectra are flux-calibrated and dereddened
for Galactic extinction, 
spectroscopic redshifts are measured through template fitting. 
For line flux measurement, 
each spectrum is first visually inspected 
to check whether the spectroscopic redshift estimation is 
reasonable enough,
then deredshifted into the rest frame using 
the spectroscopic redshift. 
Fluxes are measured for strong emission lines 
([O\,\textsc{ii}], [O\,\textsc{iii}]$\lambda\lambda$4959, 5007, 
H$\alpha$, H$\beta$,
[N\,\textsc{ii}]$\lambda\lambda$6548, 6584), 
applying a Gaussian profile fitting to the 
continuum-subtracted spectrum 
with the \textsc{mpfit}\footnote{http://purl.com/net/mpfit} 
package based on the Levenberg-Marquardt method
\citep{2009ASPC..411..251M}.
The local continuum around each emission line 
is determined by the linear fit, 
except for the hydrogen lines 
where the stellar absorption is taken into account 
as a Gaussian with a broad FWHM
and negative amplitude.
For details of the redshift estimation 
and line flux measurement, please see 
\citet{2013ApJS..207...37S}.

\subsection{Sample Selection}
\label{sec:sample}

The numbers of MIR-selected sources targeted for 
optical spectroscopic observations 
are 1155 \citep[1796, including targets selected based on 
other selection criteria; ][]{2013ApJS..207...37S}
and 1505,
in observations that are obtained by 2008 
and during the years 2020-2021, respectively. 
Since our motivation is to compare 
the gas-phase metallicity (measured in optical spectra) 
and the relative PAH strength to total dust luminosity 
(estimated from the MIR-FIR SED),
we apply the following criteria for sample selection:
(1) the object is detected 
in at least one of the Herschel/SPIRE bands 
(i.e., 250, 350, and 500\,$\mu$m)
with the flux density larger than the 10\,mJy,
(2) the object has at least four MIR
photometric points at wavelengths between 7 and 24\,$\mu$m, 
and (3) emission lines of H$\alpha$ and [N\,\textsc{ii}]$\lambda$6584 
are clearly detected (i.e., S/N$\ge3$) in the spectrum of the object.

The existence of flux density at
wavelengths longer than the rest frame $100\,\mu$m
is essential to constrain the total IR luminosity,
since those wavelengths sample beyond the peak of 
the cold dust emission inside a galaxy. 
We measure flux densities in Herschel/SPIRE 
250, 350, and 500\,$\mu$m mosaic maps 
provided by the
Herschel Extragalactic Legacy Project
\citep[][]{2021MNRAS.507..129S}
using the deblender tool \textsc{xid+} \citep{2017MNRAS.464..885H}
with the optical source coordinates as priors. 
The use of a de-blender tool at a source position 
allows measurement of fluxes of sources that are not detectable
by a blind source detection over several times the rms level. 
Nevertheless,
in principle it is unreliable to probe below the confusion limit 
of the Herschel/SPIRE \citep{2010A&A...518L...5N}.
Therefore we adopt 10\,mJy as the flux density cut in SPIRE bands, 
and assign the upper limit of 10\,mJy for nondetected objects.
Note that we have compared 250\,$\mu$m flux densities 
measured at the optical ($g$-band) source position 
and mid-infrared (9\,$\mu$m) source position, 
to find that the two are consistent,
which suggests that it is not necessary to worry about 
the spatial offset between dust and star 
for MIR sources in our sample. 

We impose a criterion that at least four
MIR photometric points are required 
in order to constrain the PAH strength. 
The MIR photometric points used are 
from the catalog of \citet{2012A&A...548A..29K}, 
i.e., the original AKARI/IRC MIR source catalog. 
The catalog is constructed by merging flux densities 
measured separately in different wavelength images,
therefore the availability of at least four photometric points 
indicates 
that the MIR detection is robust enough.  

Finally, in order to estimate the gas-phase metallicity 
using ratios between strong emission lines
(will be discussed in Section~\ref{sec:metal}),
it is required that 
at least H$\alpha$ and [N\,\textsc{ii}]$\lambda$6584 
line fluxes are available. 
For spectral classification of star-formation dominated galaxies, 
H$\beta$ and [O\,\textsc{iii}]$\lambda$5007 line fluxes
are required in addition to 
H$\alpha$ and [N\,\textsc{ii}]$\lambda$6584. 
Considering the wavelength coverage, 
this criterion limits 
the redshift of our sample galaxies to be at $z<0.4$. 

In order to use the metallicity estimators based on 
emission line ratios 
that are either empirically calibrated using star-forming galaxies
or theoretically modeled for H\,\textsc{ii} regions,
we exclude objects possibly dominated by active galactic nucleus (AGN)
using a BPT diagram (Figure~\ref{fig:bpt}a)
and a MIR color-color diagram (Figure~\ref{fig:bpt}b). 
In the BPT diagram, 
objects located below the 
AGN separation line \citep{2003MNRAS.346.1055K}
are considered to be H\,\textsc{ii} galaxies,
while the objects between the AGN line
and the extreme starburst line \citep{2001ApJ...556..121K}
are classified as composite systems \citep{2006MNRAS.372..961K}.
Objects classified into three categories 
(H\,\textsc{ii}, composite, and AGN) in the BPT diagram
(Figure~\ref{fig:bpt}a) are differently populated 
in MIR color-color diagram (Figure~\ref{fig:bpt}b), 
since MIR colors are affected by warm dusty torus around AGN. 
Almost, if not all, of the star-forming galaxies 
(``H\,\textsc{ii}'') 
are located outside the AGN selection wedge
in the MIR color-color diagram. 
Note that only 25\% of the AGN 
classified by optical spectral line ratios 
are located within the AGN selection window/wedge.
This is explained by the fact that
optical spectroscopic surveys are likely to be biased
toward unobscured AGN,
while MIR color-based AGN selection may identify 
highly obscured AGN. 
Completeness and reliability of AGN candidate selection
are known to depend on the MIR color cut and magnitude 
\citep[][]{2018ApJS..234...23A}. 
Considering MIR-selected AGN, 
the selection window/wedge in WISE color-color
diagram is expected to be about 75\% complete 
and fairly ($>80$\%) reliable 
\citep[][]{2018ApJS..234...23A}. 
Therefore, we use the selection window 
in the WISE color-color diagram 
to define photometrically selected AGN 
in case any of the four emission lines
(H$\beta$, [O\,\textsc{iii}]$\lambda5007$, H$\alpha$, and 
[N\,\textsc{ii}]$\lambda6584$) is not detected. 
We exclude spectroscopically classified AGN/composite objects 
and photometrically selected AGN 
to construct the final sample of galaxies 
for metallicity measurement
and SED fitting analysis.

The number of sample galaxies 
that satisfy
(1) number of FIR (250, 350, and 500\,$\mu$m) photometry,
(2) number of MIR (7-24\,$\mu$m) photometry, 
(3) H$\alpha$ and [N\,\textsc{ii}] line detection 
as well as $z_\mathrm{spec} < 0.4$,
and do not meet any of the AGN selection condition 
is 373. 
These galaxies are discussed in Section~\ref{sec:result}.

\section{Metallicity Estimation}    
\label{sec:metal}

\begin{figure*}
\epsscale{1.1}
\plottwo{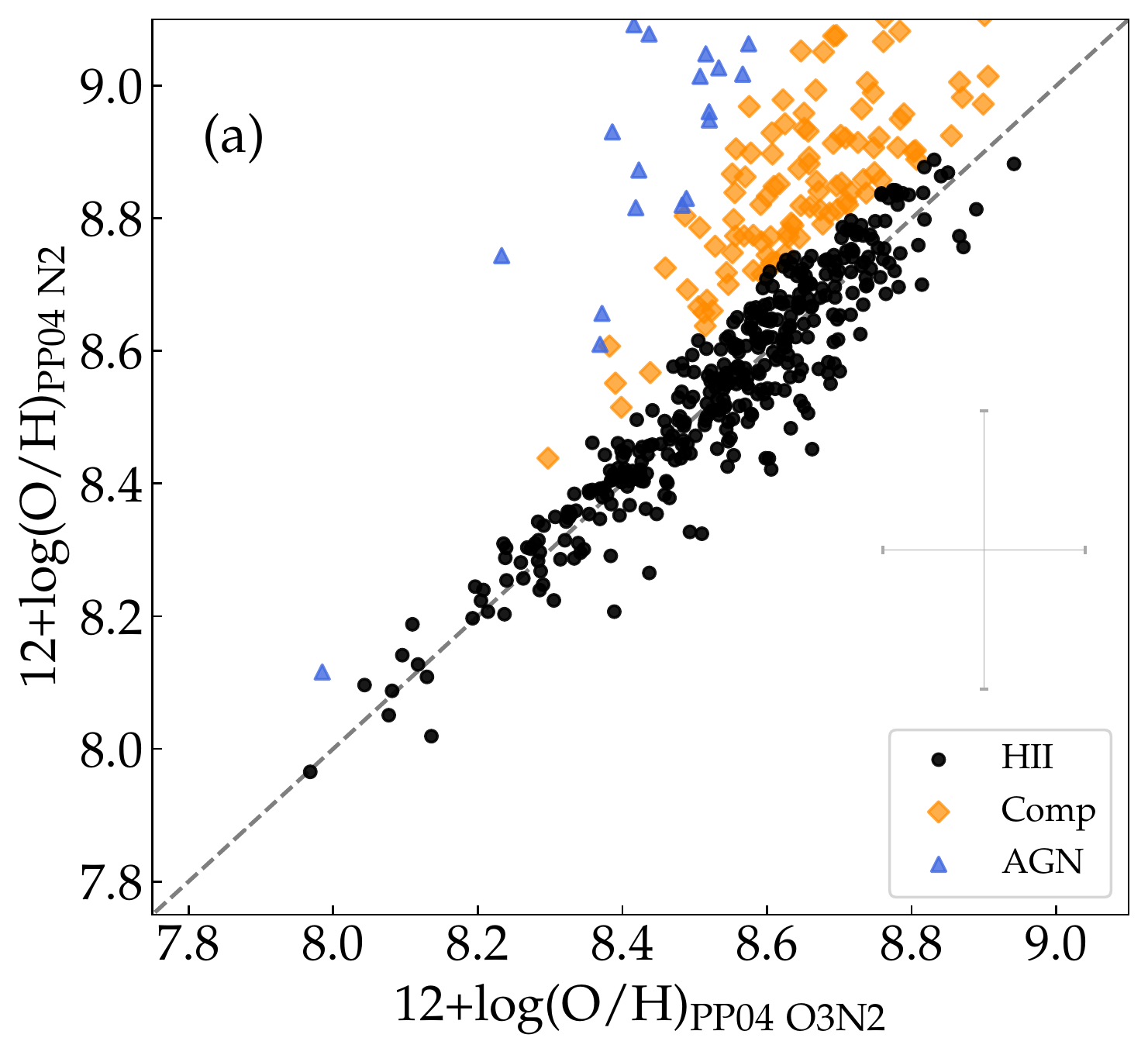}{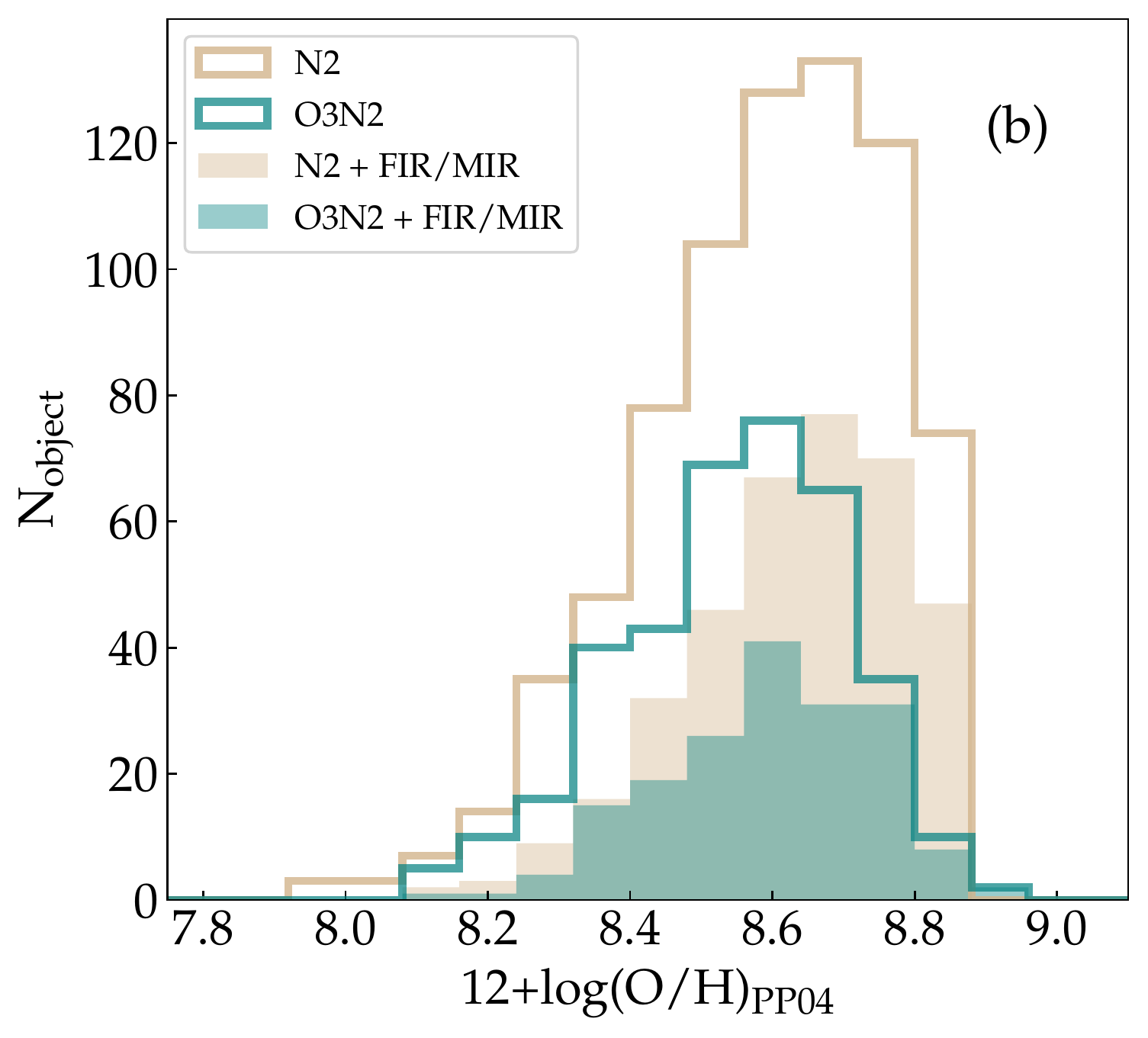}
\caption{\label{fig:metal}
        (a) Comparison between the oxygen abundance values
        derived using O3N2 index and N2 index \citep{2004MNRAS.348L..59P}.
        The uncertainties of $12+\mathrm{log(O/H)}$ implemented in two methods
        are 0.14\,dex and 0.18\,dex (at $1\sigma$)
        for O3N2 and N2, respectively. Such systematic uncertainty
        and uncertainty in the measured line flux ratio
        are quadratically summed to produce a typical error bar,
        marked in the corner.
        For H\textsc{ii} galaxies,
        $12+\mathrm{log(O/H)}$ values from O3N2 and N2
        are consistent within 0.02.
        (b) Distribution of the $12+\mathrm{log(O/H)}$ values
        for spectroscopically observed objects
        (after the AGN exclusion),
        estimated using O3N2 index (371 objects)
        and N2 index (747 objects).
        Filled histograms
        are the metallicity distributions of
        objects satisfying criteria for
        availability of FIR and MIR photometric points
        (179 and 369 for O3N2 and N2 methods, respectively;
        the number of total objects with measured metallicity
        that meets other criteria is 373).
}
\end{figure*}

Among different metallicity indicators, 
one of the most widely used is the N2 index 
($\mathrm{N2}\equiv\mathrm{log}{([\mathrm{N}\,\textsc{ii}]\lambda6584/\mathrm{H}\alpha})$),
while the ionization field strength that determines
the observed nitrogen line flux 
is dependent on the gas-phase metallicity.
In the low-metallicity 
(less than 20\,\% solar metallicity)
and high-metallicity (above solar) regime,
the higher-order polynomial fit is suggested
to describe the relationship between N2 and $12+\mathrm{log(O/H)}$,
rather than forcing a linear fit. 
\citet{2004MNRAS.348L..59P} have suggested
that the following equation is
valid for extragalactic H\,\textsc{ii} regions
in the range $-2.5 < \mathrm{N2} < -0.3$,
with $1\sigma$ uncertainty of 0.18 dex:

\begin{equation}
\label{eqn:PP04-N2}
12 + \mathrm{log(O/H)} = 9.37 + 2.03\mathrm{N2} + 1.26\mathrm{N2}^2 + 0.32\mathrm{N2}^3. 
\end{equation}

The merit of using the N2 index is that 
the uncertainty in extinction correction to derive this index is small 
since the [N\,\textsc{ii}]$\lambda6584$ and H$\alpha$
lines are closely located in terms of the wavelength. 
The method is also applicable to the cases 
when no lines are detected in shorter wavelengths.
However, in high-metallicity regime
where [N\,\textsc{ii}] line saturates, 
adding [O\,\textsc{iii}] line 
is considered to provide
better constraint on the metallicity 
as the [O\,\textsc{iii}] keeps decreasing 
with a decrease in oxygen abundance. 
An empirical calibration using 
[N\,\textsc{ii}]/H$\alpha$ and [O\,\textsc{iii}]/H$\beta$ line ratios, 
measured in the observed optical spectra, 
is known to be the O3N2 method. 
The O3N2 index is defined as
$\mathrm{O3N2}\equiv\mathrm{log}${(([O\,\textsc{iii}]$\lambda5007$/H$\beta$)/([N\,\textsc{ii}]$\lambda6584$/H$\alpha$))}.
Based on the good linear correlation 
between the O3N2 and oxygen abundance
of individual extragalactic H\,\textsc{ii} regions, 
the following conversion formula is suggested by 
\citet[][hereafter PP04 O3N2]{2004MNRAS.348L..59P}:

\begin{equation}
	\label{eqn:PP04-O3N2}
	12 + \mathrm{log(O/H)} = 8.73 - 0.32\,\mathrm{O3N2}
\end{equation}

\noindent of which $1\sigma$ systematic uncertainty is 0.14 dex.
This equation is valid in the range $\mathrm{O3N2}<2$,
corresponding to environments above 0.4 solar metallicity.

Since the uncertainty in the calibration formula
is smaller in case of the O3N2 method than the N2 method, 
we use Equation~\ref{eqn:PP04-O3N2}
to derive gas-phase metallicity of sample galaxies.
If [O\,\textsc{iii}] and/or H$\beta$ line 
are not detected in the optical spectrum
(i.e., $\mathrm{S/N} < 3$), 
Equation~\ref{eqn:PP04-N2} is used instead. 
To derive line flux ratios, 
the measured line fluxes are
corrected for dust attenuation
using standard Milky Way reddening curve \citep{1958AJ.....63..201W}
and the observed Balmer decrement H$\alpha$/H$\beta$
if both H$\alpha$ and H$\beta$ line fluxes are available. 
Recent work \citep{2022NatAs...6..844C} has warned that
the use of rest-frame optical lines
may underpredict the oxygen abundance
for heavily obscured galaxies such as ultraluminous infrared galaxies
(ULIRGs).
Since the total IR luminosity of our sample galaxies 
is mostly $\mathrm{log} L_\mathrm{IR} [\mathrm{L}_\odot] < 11.5$,   
none of our sample galaxies are classified as ULIRGs. 
The median of the extinction estimates, 
$\langle A_V\rangle$, of the sample galaxies  
are calculated to be 1.4\,mag based on the H$\alpha$/H$\beta$ ratio.
This value is comparable to that of
local star-forming galaxies from SDSS 
\citep[e.g.,][]{2003ApJ...599..971H}, 
and is significantly smaller than 
$A_V (\mathrm{H}\alpha/\mathrm{H}\beta)\simeq2.5$-2.9\,mag 
of local ULIRGs
and high-redshift submillimeter galaxies 
\citep[e.g.,][]{2004A&A...415..885F, 2006ApJ...651..713T}. 
Therefore we expect that 
it is appropriate to use
the extinction-corrected optical line ratios
in calculation of the oxygen abundance 
for our spectroscopic sample of MIR-selected galaxies.

Figure~\ref{fig:metal}a shows that 
the oxygen abundance values ($12+\mathrm{log(O/H)}$)
derived using O3N2 and N2 
are consistent with each other 
with a scatter of less than 0.02 dex 
in case of H\textsc{ii} galaxies.
Objects for which metallicity is estimated using N2 index
since O3N2 index is not available
are more attenuated (about twice larger Balmer decrement)
and slightly less luminous in H$\alpha$
(factor of 1.5 lower mean H$\alpha$ flux)
compared to objects for which O3N2 index is available.
Figure~\ref{fig:metal}b shows the metallicity distribution
of non-AGN objects.
The metallicity 
of star-formation dominated galaxies 
ranges $7.9 < 12+\mathrm{log(O/H)} < 9.0$, 
while the median metallicity is close to that of solar.
If the sample criteria for FIR and MIR photometric data points 
are applied, 
the median metallicity for such galaxies
is slightly higher.
This suggests that our sample galaxies are 
likely to be affiliated in the massive galaxies
among the entire MIR-selected galaxies, 
which is natural enough considering the FIR and MIR detection limits. 
As is already shown in Figure~\ref{fig:specsample}c, 
the final sample galaxies with MIR/FIR detection 
(marked in orrange) 
show slightly larger median stellar mass 
($\langle\mathrm{log}\,M_\mathrm{star}[\mathrm{M}_\odot]\rangle=10.6$)
and SFR 
($\langle\mathrm{log}\,\mathrm{SFR}[\mathrm{M}_\odot\,\mathrm{yr}^{-1}]\rangle=0.3$)
compared to entire spectroscopic sample. 
Therefore, our sample galaxies
used in the investigation of metallicity-PAH relation are
star-forming galaxies that lie above the median SFR
of MIR-selected population.
The SED fitting procedure to estimate physical parameters
such as stellar mass and SFR
is discussed in Section~\ref{sec:sedfit}.

\begin{figure}
\epsscale{1.1}
\plotone{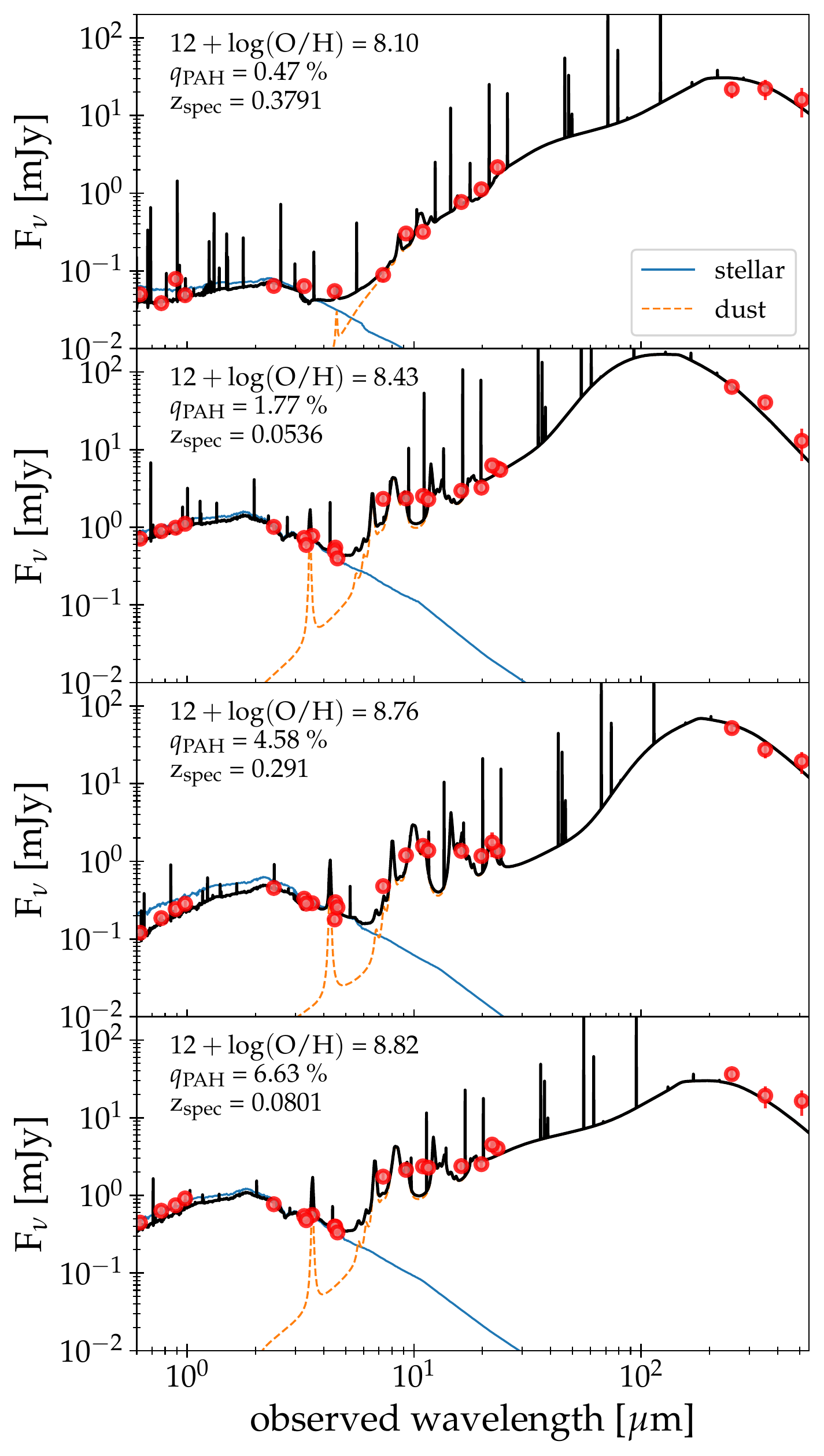}
\caption{\label{fig:sed}
        Example MIR-FIR SED of MIR-selected galaxies
        (best-fit template from \textsc{cigale} run).
        Blue solid and orange dashed lines represent
        (unattenuated) stellar component and
        dust emission component, respectively.
	Oxygen abundance $12+\mathrm{log(O/H)}$,
	PAH mass fraction derived from SED fitting,
	and redshift 
	are indicated.
        }
\end{figure}

\begin{table*}
	\caption{Parameters for SED Fitting Using \textsc{cigale}}
	\label{tab:cigale_params}
	\begin{tabular}{ll}
		\hline
		Model and Input parameters & Range \\ 
		\hline
		\hline
		Star-formation history: \texttt{sfhdelayed} \\ 
		\hline
		\hspace{4pt} e-folding time of the main stellar population model [Myr] & 1000, 3000, 15000 \\
		\hspace{4pt} Age of the main stellar population in the galaxy [Myr] & 500, 2000, 5000, 6000, 7000, 8000, 9000, 10,000, 12,000 \\
		\hspace{4pt} e-folding time of the late starburst population [Myr] & 50 \\
		\hspace{4pt} Age of the late starburst population [Myr] & 20, 50, 100 \\
		\hspace{4pt} Mass fraction of the late burst population & 0.0, 0.1, 0.2, 0.3, 0.5 \\ 
		\hline
		Stellar population: \texttt{bc03} \\ 
		\hline
		\hspace{4pt} Initial mass function & Chabrier \\ 
		\hspace{4pt} Metallicity & 0.02 \\
		\hline
		Dust attenuation: \texttt{dustatt\_modified\_CF00} \\ 
		\hline
		\hspace{4pt} Av\_ISM & 0.3, 0.6, 1.0, 1.6, 2.3, 3.8 \\
		\hspace{4pt} Power-law slope of the attenuation in the ISM & $-$0.7 \\
		\hspace{4pt} Power-law slope of the attenuation in the birth clouds & $-$1.3 \\
		\hline
		Dust emission: \texttt{dl2014} \\ 
		\hline
		\hspace{4pt} Mass fraction of PAH ($q_\mathrm{PAH}$) & 0.47, 1.12, 1.77, 2.50, 3.19, 3.90, 4.58, 5.26, 5.95, 6.63, 7.32 \\
		\hspace{4pt} Minimum radiation field ($U_\mathrm{min}$) & 0.20, 1.00, 5.00, 10.00, 15.00, 25.00, 40.00 \\
		\hspace{4pt} Power-law slope index ${\tilde\alpha}$ ($dU/dM \propto U^{\tilde\alpha}$) & 1.6, 2.0, 2.3, 2.5, 2.7, 3.0 \\
		\hspace{4pt} Fraction illuminated from $U_{\rm min}$ to $U_{\rm max}$ ($\gamma$) & 0.00, 0.01, 0.02, 0.05, 0.1, 0.2, 0.3, 0.4, 0.5 \\
		\hline
	\end{tabular}
\end{table*}

Note that in \citet{2004MNRAS.348L..59P},
the number of high metallicity H\,\textsc{ii} regions
($12+\mathrm{log(O/H)}\simeq9.0$, $\mathrm{O3N2}<0$)
is small
and oxygen abundances of such H\,\textsc{ii} regions
are deduced from photoionization models
rather than the electron temperature ($T_e$) method.
\citet{2013A&A...559A.114M} have suggested
a slightly flatter correlation between
$12+\mathrm{log(O/H)}$ and O3N2
using the increased number of extragalactic H\,\textsc{ii} regions
with $T_e$-based abundance measurements.
The two calibrations agree with each other
in low metallicity ranges,
yet the difference between the two increases
as the metallicity increases. For example,
with $\mathrm{O3N2}=-1$,
$12+\mathrm{log(O/H)}$ values calculated using
\citet{2004MNRAS.348L..59P}
and \citet{2013A&A...559A.114M}
are 9.0 and 8.7, respectively.
The N2 methods suggested in \citet{2004MNRAS.348L..59P}
and \citet{2013A&A...559A.114M}
have similar trend that \citet{2004MNRAS.348L..59P}
formula leads to higher metallicity
than that of \citet{2013A&A...559A.114M}.
We use the N2 and O3N2 method of
\citet{2004MNRAS.348L..59P} in this study,
and relatively metal-poor and metal-rich galaxies 
can be distinguished 
with a consistent manner. 
However, it should be noted that
the derived $12+\mathrm{log(O/H)}$ values
may not be the absolute oxygen abundance
due to calibration uncertainty,
and such uncertainty should be taken into account 
when comparing different works based on 
different methods.

\section{SED Fitting}   \label{sec:sedfit}

We fit the observed photometric points
to the modeled 
SEDs 
using \textsc{cigale}\footnote{https://cigale.lam.fr/2022/07/04/version-2022-1/}
\citep[Code Investigating GAlaxy Emission,][]{2009A&A...507.1793N, 2019A&A...622A.103B}
to derive physical parameters of galaxies. 
The code \textsc{cigale} enables us to remove
the stellar contribution from the observed SED 
by computing spectral template of a galaxy 
that consists of both stellar and dust component 
based on an energy balance principle, in which
the energy from stars is reprocessed by dust 
to produce infrared emission.
Stellar population synthesis model 
is convolved with the chosen star-formation history, 
which is then attenuated by dust.
The computed spectral model
(i.e., the sum of stellar component and dust emission component) 
is compared to the observed photometric points 
(or flux upper limits in case of nondetection)
from UV to FIR wavelengths (0.1-500\,$\mu$m). 
Figure~\ref{fig:sed} shows examples of best-fit templates
with observed photometric data points
for representative galaxies
with different metallicity values
(calculated in Section~\ref{sec:metal}).
The input parameters we use to run the modules of \textsc{cigale}
are summarized in Table~\ref{tab:cigale_params}.

To calculate the stellar component, 
the ``delayed'' star-formation history model
(\texttt{sfhdelayed}) is used. 
In this model, the SFR 
changes relatively smoothly according to the following formula: 

\begin{equation}
\mathrm{SFR}(t) \propto \frac{t}{\tau^2} ~\mathrm{exp}\left(-\frac{t}{\tau} \right),
\end{equation}

\noindent which is valid at $0<t<t_0$, where $t_0$
is the time of star formation onset
and $\tau$ is the time of SFR peak.
By using small $\tau$ and large $\tau$,
nearby passive galaxies and actively star-forming galaxies are
efficiently modeled without the introduction of
an abrupt, extreme change in SFR.
For estimation of SFR and stellar mass,
we choose to use the stellar population synthesis model of
\citet{2003MNRAS.344.1000B} and \citet{2003PASP..115..763C}
initial mass function.
The stellar population metallicity is fixed to solar (0.02),
since the 
relationship between oxygen abundance (gas metallicity) 
and stellar metallicity is not straightforward. 
The stellar component is reddened 
using the 
extended \citet{2000ApJ...539..718C}
attenuation law, 
by allowing the attenuation $A_V$ 
to vary between 0.3 and 3.8 mag.

\begin{figure*}
\epsscale{1.1}
\plotone{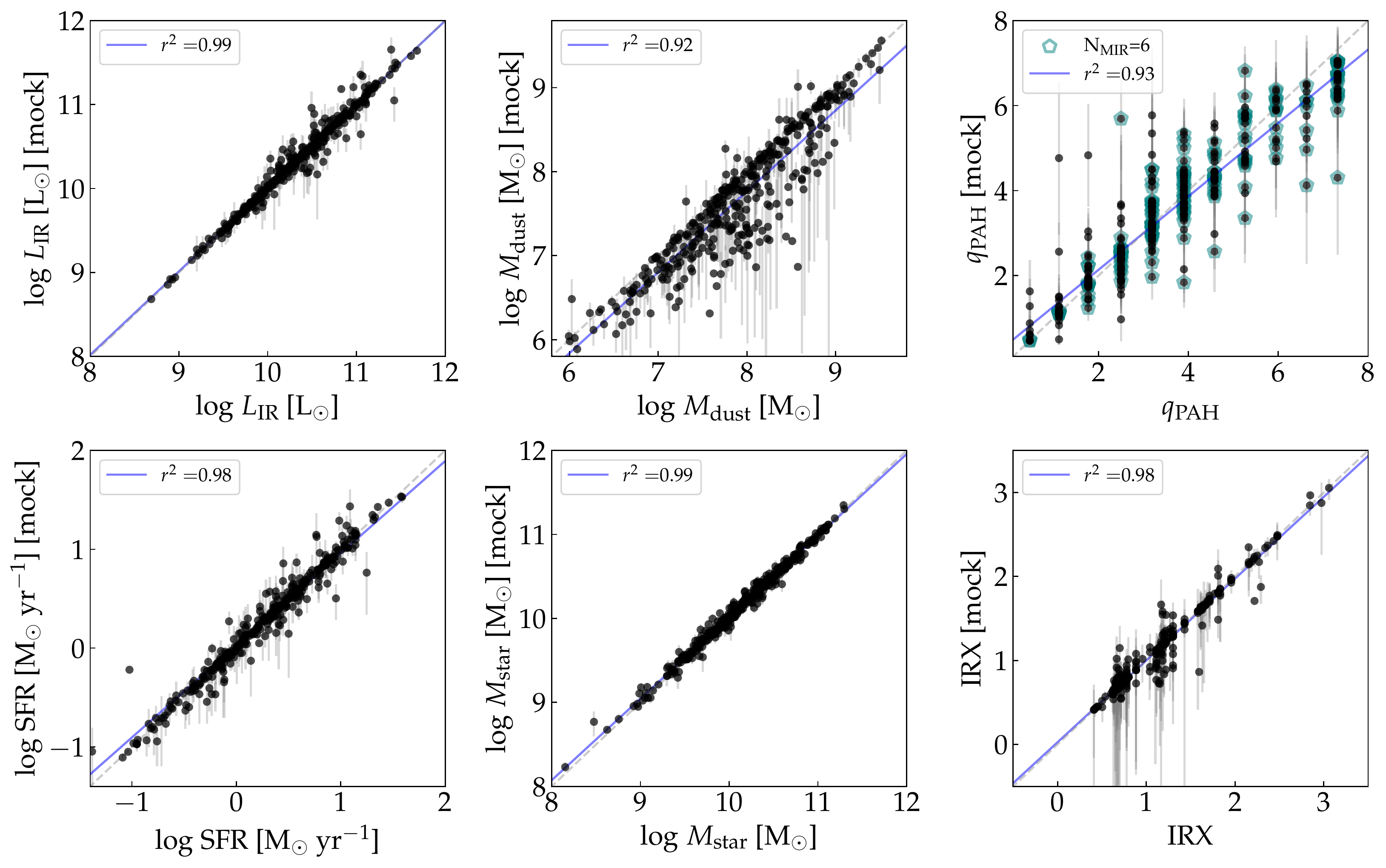}
\caption{\label{fig:mock}
        Test results to reproduce values of physical parameters
        based on the \textsc{cigale} run using mock catalog
        constructed from the best-fit SEDs.
        373 non-AGN objects with estimated metallicity values
        are used to construct these plots.
        The $x$-axis represents the total IR luminosity,
        dust mass, mass fraction of PAH particles ($q_\mathrm{PAH}$),
        SFR, stellar mass, and IR excess (defined as
        the ratio between IR and UV luminosity)
        determined from the best-fit SED.
        A mock catalog is constructed
        using the expected flux density
        (based on the best-fit SED)
        and the observed flux uncertainty in each filter,
        which is to be used as an input
        in the \textsc{cigale} run with the same configuration
        to derive output ($y$-axis) value for each physical parameter.
        Gray dashed lines are $y=x$ lines,
        while the correlations between the input and output values
        are described as linear functions (blue solid lines)
        with the determination coefficients $r^2$.
        All properties are relatively well reproduced
        with strong correlation ($r^2 > 0.9$),
        supporting the robustness of the derived properties.
}
\end{figure*}

The key module to calculate the MIR-FIR SED is
a dust emission module.
\textsc{cigale} provides several ways of describing
dust emission: semiempirical template \citep{2014ApJ...784...83D};
two-component model \citep{2007ApJ...657..810D, 2014ApJ...780..172D};
and the analytic model \citep{2012MNRAS.425.3094C},
as well as the model considering dust evolution
according to local ISM physical condition \citep[THEMIS;][]{2017A&A...602A..46J}.
The dust templates of \citet{2014ApJ...784...83D}
are constructed based on the modified SEDs
of nearby star-forming galaxies
with the addition of the AGN component;
however,
the model allows only limited variation in PAH emission.
The \citet{2012MNRAS.425.3094C} model
consists of a single-temperature modified blackbody emission
dominating FIR wavelengths
and power-law continuum in MIR wavelengths,
i.e., not taking PAH emission into account.
The THEMIS model
uses the fraction of
HAC (hydrogenated amorphous carbons)
as free parameter instead of PAH fraction,
although the two are in general proportional to each other.
Therefore, in order to estimate the PAH contribution
in the SEDs of galaxies,
we use the \texttt{dl2014} module for modeling dust emission
that is a refined version of the \citet{2007ApJ...657..810D} model.

In \citet{2007ApJ...657..810D} dust emission model, 
the spectral templates are calculated
assuming that dust (silicate and graphite grains 
including various PAH particles) are heated by starlight. 
The starlight intensity $U$ 
(often characterized using minimum or maximum values,
and the power-law slope $\tilde\alpha$ between them), 
the fraction of dust mass 
exposed to a starlight intensity distribution $\gamma$, 
and the PAH mass fraction $q_\mathrm{PAH}$ 
are free parameters for the spectral template calculation. 
Possible $q_\mathrm{PAH}$ values are 
[0.47, 1.12, 1.77, 2.50, 3.19, 3.90, 4.58] (\%), 
which are tweaked to reproduce the Milky way extinction curve. 
\citet{2014ApJ...780..172D} 
have presented a slightly revised version of physical dust model of 
\citet{2007ApJ...657..810D}, 
allowing a larger range of $q_\mathrm{PAH}$
-- [0.47, 1.12, 1.77, 2.50, 3.19, 3.90, 4.58, 5.26, 5.95, 6.63, 7.32] -- 
with slightly different normalization in total dust mass estimation. 
We use a wide range of $q_\mathrm{PAH}$, 
$U_\mathrm{min}$, $\tilde\alpha$, 
and $\gamma$ (Table~\ref{tab:cigale_params}) 
to reproduce diversities of MIR-FIR SED shape. 

In addition to the modules and parameters
in Table~\ref{tab:cigale_params},
we include nebular emission (\texttt{nebular})
and AGN (\texttt{fritz2006}) modules to compute models.
However, nebular emission model parameters
are fixed to default values
(fixed ionization parameter, electron density
and no Lyman continuum escape),
and AGN module is practically switched off
by fixing the AGN fraction to be zero.
The reason for not considering the AGN component is
because AGN-dominated galaxies,
selected either spectroscopically or photometrically,
are excluded from the sample (Figures~\ref{fig:bpt}a and \ref{fig:bpt}b).
To test whether such a strategy is reasonable,
we fit SEDs of
179 spectroscopically selected H\,\textsc{ii} galaxies
(for which O3N2 metallicity measurement is available)
by switching the AGN module on,
then compared best-fit models
to that derived without the AGN component.
The AGN component is calculated
by varying AGN fraction
(the ratio of AGN luminosity to total luminosity)
to be between 0 and 1, with the range of
9.7\,$\mu$m optical depth to be 0.6--6.0,
and allowing opening angle between type 1 and type 2 AGN.
For only two out of 179 (about 1\,\%) objects,
the best-fit model is different,
where the inclusion of the AGN component leads to
a model with smaller $\chi^2$.
The parameters associated with the AGN module
indicate that these two prefer the AGN component
contributing $\sim10$\,\% of the total IR luminosity
with a low optical depth of 9.7\,$\mu$m silicate absorption.
Considering the small number of such cases,
the AGN fraction is assumed to be negligible
in our sample of galaxies.

Physical properties of galaxies, such as the SFR, 
mass of the stellar and dust component, dust attenuation, 
dust luminosity (i.e., total IR luminosity), 
and PAH mass fraction 
are derived from the \textsc{cigale} run. 
Instead of extracting physical parameter values 
from the best-fit model, 
\textsc{cigale} probes 
the possible range for each parameter
by taking the mean and standard deviation 
of the probability distribution function 
weighted for the goodness of fit. 
Such a strategy provides more robust estimates of properties 
and an idea of their uncertainties 
in case there exists degeneracy between physical properties. 
Based on this, it is also possible 
to investigate the reliability of the derived values 
of physical properties,
through the ``mock test'' provided by \textsc{cigale}.
In the mock test, once the best-fit SED model is determined for a galaxy, 
the synthetic flux density in each filter 
can be calculated from the best-fit model. 
By running the \textsc{cigale} again 
using the calculated synthetic flux densities 
and the observed flux uncertainties (i.e., mock catalog)  
with exactly the same configuration as the original run, 
it is possible to check whether the input physical values 
(values from the best-fit model) 
are well recovered, 
which allows us to assess the reliability of the derived values. 

\begin{figure*}
\epsscale{1.15}
\plotone{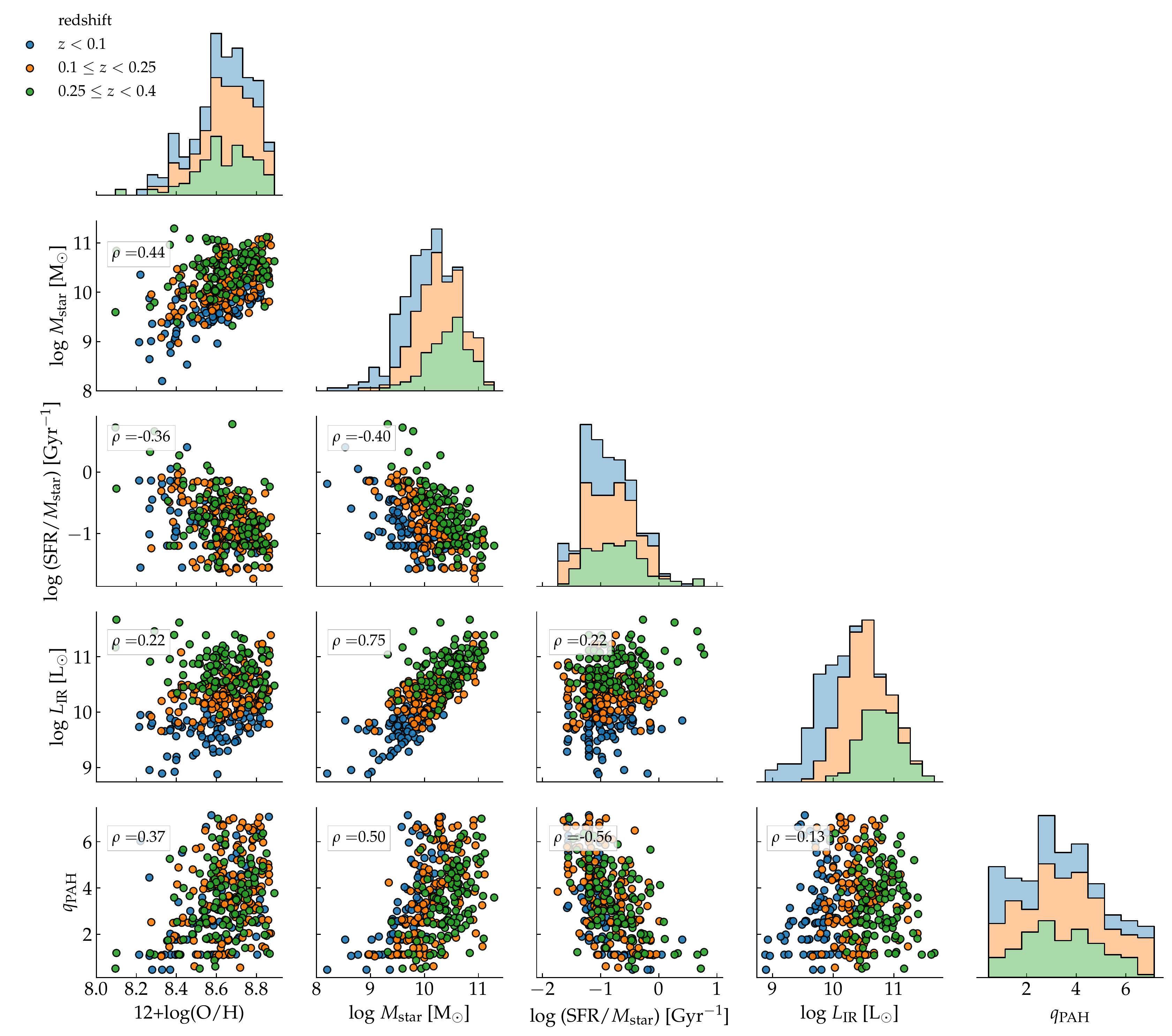}
        \caption{\label{fig:figall}
        Correlation between different physical parameters:
        gas-phase metallicity $12+\mathrm{log(O/H)}$,
        stellar mass, specific SFR (sSFR),
        total IR luminosity,
        and mass fraction of PAH particles ($q_\mathrm{PAH}$).
        Spearman rank coefficients ($\rho$) between two variables
        in the $x$- and $y$-axis are specified in each panel.
        The sample
        is divided into three subsamples
        according to the redshift ($z<0.1$, $0.1\le z < 0.25$,
        and $0.25\le z <0.4$),
        which results in a similar number of galaxies
        in each subsample.
        The distribution of each physical parameter
        is illustrated using a stacked histogram
        (in order of $0.25\le z <0.4$, $0.1\le z < 0.25$,
        and $z<0.1$ bins)
        in order to better represent
        the difference between different subsamples.
        }
\end{figure*}

The results of the mock test are summarized 
in Figure~\ref{fig:mock}. 
Values for total IR luminosity, SFR, stellar mass, 
and IR excess (ratio between IR luminosity and UV luminosity) 
are robustly reproduced 
by a repeated \textsc{cigale} run
that accounts for the observed flux uncertainties.
In contrast to IR luminosity and stellar mass, 
the scatter between the input and output values 
is relatively large for the total dust mass 
and the PAH mass fraction. 
However, 
because the correlation coefficient is large enough 
($r^2 > 0.9$), 
the relative comparison of $q_\mathrm{PAH}$ values 
between different galaxies is still valid. 
In panel for $q_\mathrm{PAH}$,
galaxies that are detected in all six bands of 
the AKARI/IRC MIR wavelengths 
are separately marked.
Although applying a criterion 
for number of MIR photometric points 
($N_\mathrm{MIR}$) to be six 
removes some galaxies with large uncertainty in $q_\mathrm{PAH}$, 
it is not clear that the correlation coefficient 
becomes larger with $N_\mathrm{MIR}=6$
compared to the case of $N_\mathrm{MIR}\ge4$.
Therefore, the criterion is kept to be 
$N_\mathrm{MIR}\ge4$ 
as mentioned in Section~\ref{sec:sample}.

\section{Results}   \label{sec:result}

\begin{figure*}
\epsscale{1.15}
\plotone{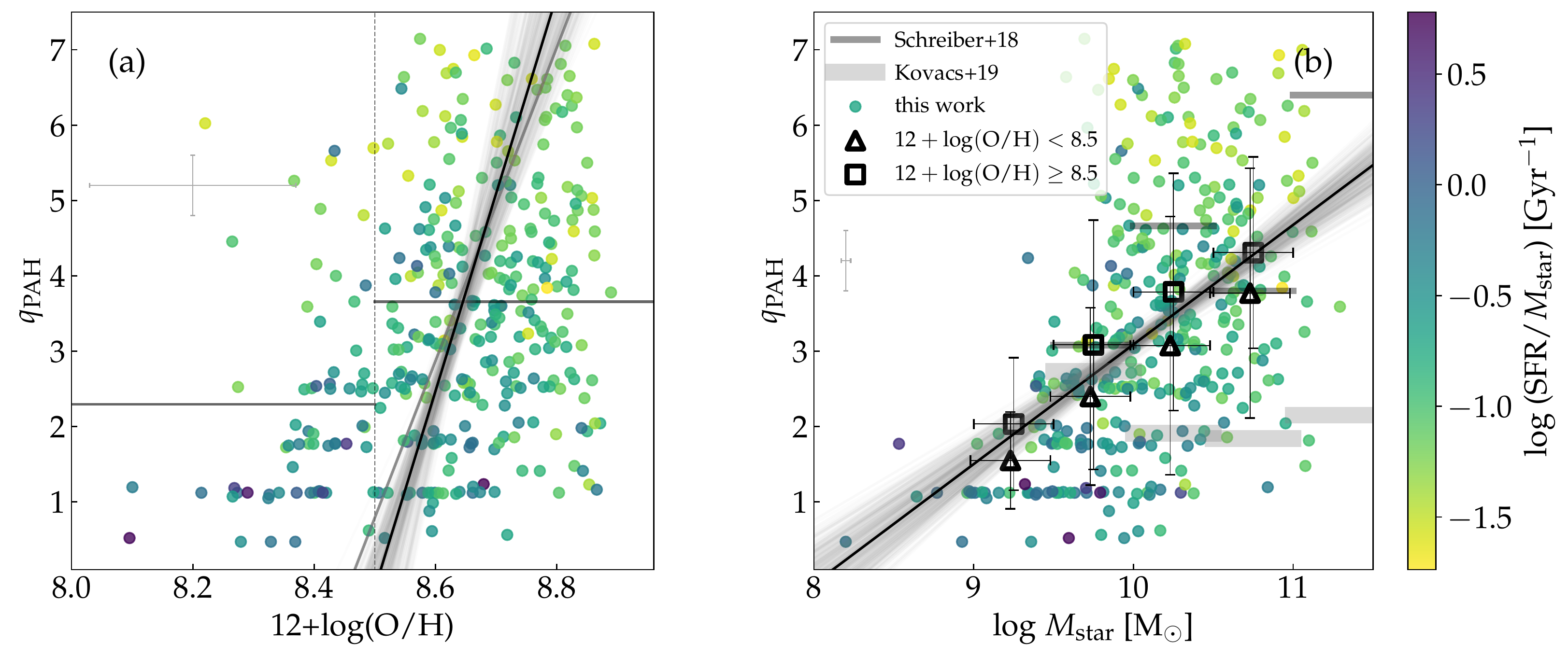}
\caption{\label{fig:qpah}
        (a) $q_\mathrm{PAH}$ vs. $12+\mathrm{log(O/H)}$
        for 373 non-AGN MIR-selected galaxies
        (same plot as the panel (4,0)
        in Figure~\ref{fig:figall}).
        Symbols are color-coded according to the
        sSFR. Typical error bar in oxygen abundance
        and $q_\mathrm{PAH}$ is marked in the corner.
        The mean value of the $q_\mathrm{PAH}$
        is 2.3 and 3.7
        for galaxies with $12+\mathrm{log(O/H)}<8.5$
        and $12+\mathrm{log(O/H)}\ge8.5$, respectively
        (horizontal lines).
        The black line is the linear regression
        describing the relationship between $12+\mathrm{log(O/H)}$
        and $q_\mathrm{PAH}$,
        while gray lines are possibilities of linear correlation
        determined based on Bayesian methods
        considering uncertainties in $12+\mathrm{log(O/H)}$
        and $q_\mathrm{PAH}$.
        (b) $q_\mathrm{PAH}$ vs. stellar mass.
        For galaxies in stellar mass bins of
        $\mathrm{log}(M_\mathrm{star}/\mathrm{M}_\odot)=$[9.0, 9.5],
        [9.5, 10.0], [10.0, 10.5], and [10.5, 11.0],
        galaxies are divided into the low metallicity
        ($12+\mathrm{log(O/H)}<8.5$)
        and high metallicity ($12+\mathrm{log(O/H)}\ge8.5$)
        subgroups
        and the mean $q_\mathrm{PAH}$
        for each subgroup are plotted
        (note that points for the low metallicity
        subgroups are slightly shifted in $x$-direction
        for clarity).
        Overplotted are (1) $q_\mathrm{PAH}$ values
        measured from the stacked SEDs
        of $0.3<z<0.7$ star-forming galaxies
        in different stellar mass bins
        \citep[black horizontal lines;][]{2018A&A...609A..30S},
        and (2) $\langle q_\mathrm{PAH}\rangle$ values
        for different stellar mass bins of $z<0.7$
        MIR/FIR-selected galaxies
        \citep[gray horizontal lines;][]{2019PASJ...71...27K}.
        The linear relationship between the $\mathrm{log}(M_\mathrm{star})$
        and $q_\mathrm{PAH}$ and its posterior distribution
        is overplotted as black and gray lines.
        }
\end{figure*}

\subsection{Correlation between Physical Parameters}  \label{sec:result1}

The typical values for physical properties
of 373 MIR-selected non-AGN galaxies 
derived through SED fitting are as follows: 
stellar mass ranging $10^{8-11}\,\mathrm{M}_\odot$,
having moderate
($\sim0.1\,\mathrm{M}_\odot\,\mathrm{yr}^{-1}$)
to large ($>10\,\mathrm{M}_\odot\,\mathrm{yr}^{-1}$) 
SFR represented by total IR luminosity of 
$10^{9-12}\,\mathrm{L}_\odot$.

Figure~\ref{fig:figall} is a scatter plot matrix 
showing correlations
between different physical parameters --  
$12+\mathrm{log(O/H)}$, stellar mass, specific SFR 
(sSFR; SFR divided by the stellar mass), 
total IR luminosity, 
and the PAH mass fraction,
as well as the distribution of each parameter 
(in the diagonal locations).
We present galaxies in different redshift bins
with different colors.
Number of galaxies in each redshift bin is 
102, 158, and 113 for 
$z<0.1$, $0.1\le z<0.25$, and $0.25\le z<0.4$ bin, respectively. 
In case of stellar mass and IR luminosity,
it is evident 
from the distribution plots 
that the properties are dependent on the 
redshift of galaxies. At higher redshifts, 
the MIR-selected sample consists of
galaxies with larger median stellar mass 
and higher median IR luminosity. 
However, in case of $q_\mathrm{PAH}$ 
and $12+\mathrm{log(O/H)}$, 
the distributions look qualitatively similar
even at different redshifts.

The Spearman's rank correlation coefficients $\rho$
between $q_\mathrm{PAH}$ 
and other parameters are 
$0.37$, $0.50$, $-0.56$, and $0.13$ 
for $12+\mathrm{log(O/H)}$, 
$\mathrm{log} M_\mathrm{star}$, 
$\mathrm{log} (\mathrm{SFR}/M_\mathrm{star})$,
and $\mathrm{log} L_\mathrm{IR}$. 
A positive correlation between $q_\mathrm{PAH}$
and gas-phase metallicity is found  
with $p$-value less than $10^{-5}$,
suggesting that PAH features are on average
weaker in lower metallicity galaxies
than in higher metallicity galaxies.
In addition, PAH mass fraction is 
positively correlated with stellar mass
and negatively correlated with sSFR
-- that is, $q_\mathrm{PAH}$ is low in low-mass galaxies 
and in galaxies with large ongoing star-formation activity.
The positive trend of increasing $q_\mathrm{PAH}$ 
as the increase of metallicity and stellar mass 
reflects the mass-metallicity relation
\citep[e.g.,][]{2004ApJ...613..898T},
which is also observed with $\rho=0.44$ in our sample. 
Although a mass-SFR correlation 
of star-forming galaxies is present as is seen in the 
$\mathrm{log} L_\mathrm{IR}$ versus $\mathrm{log} M_\mathrm{star}$ plot
(with $\rho=0.75$),
the dependence of $q_\mathrm{PAH}$ 
on the IR luminosity is relatively weak 
unlike the other physical parameters. 

\subsection{Metallicity and $q_\mathrm{PAH}$ Variation}  
\label{sec:result2}

Based on Figure~\ref{fig:figall}, 
$q_\mathrm{PAH}$ is positively correlated with 
both metallicity and stellar mass. 
In Figures~\ref{fig:qpah}a and \ref{fig:qpah}b, 
we show the linear regression results 
between dependent variable $q_\mathrm{PAH}$ 
and independent variables metallicity and stellar mass. 
When performing linear regression,
the uncertainties in the derived $12+\mathrm{log(O/H)}$, 
stellar mass,
and $q_\mathrm{PAH}$ 
are taken into account using a Bayesian approach
\citep[][]{2007ApJ...665.1489K}.
The principle of the method described in 
\citet{2007ApJ...665.1489K} 
is that with measurement errors in $x$ and $y$,
linear regression is performed for 
new independent ($\xi$) and dependent ($\eta$) variables 
that are defined as
$x_i\equiv \xi_i+\mathrm{error}(x_i)$ 
and $y_i\equiv \eta_i+\mathrm{error}(y_i)$. 
By applying the MCMC algorithm, the posterior distribution of 
the regression coefficients are derived,
along with the Pearson linear coefficient for ($x, y$).
Since the uncertainties in the $12+\mathrm{log(O/H)}$
are much larger 
than those in the stellar mass,
the linear correlation coefficient 
for $q_\mathrm{PAH}$-metallicity relation ($0.97\pm0.03$)
is higher than that of the $q_\mathrm{PAH}$-stellar mass relation
($0.52\pm0.04$).
In addition, 
the $q_\mathrm{PAH}$ values for our sample galaxies
are compared to results from previous works
\citep{2018A&A...609A..30S, 2019PASJ...71...27K},
which presented $q_\mathrm{PAH}$ according to stellar mass
(Figure~\ref{fig:qpah}b). 
Although our sample consists of less massive galaxies 
than those studied in previous works, 
$q_\mathrm{PAH}$ values are 
similar in the stellar mass range 
that overlaps with that of \citet{2018A&A...609A..30S}.
The black lines in Figures~\ref{fig:qpah}a and b are 
as follows: 

\begin{equation}
	q_\mathrm{PAH}=(-240.5\pm45.5)+(28.3\pm5.3)(\mathrm{12+log(O/H)}),
\end{equation}

\begin{equation}
	q_\mathrm{PAH}=(-12.8\pm1.5)+(1.6\pm0.2)~\mathrm{log}\,M_\mathrm{star} [\mathrm{M}_\odot].
\end{equation}

In order to investigate the variation of 
PAH strength in terms of metallicity,
in a situation where other physical parameters 
such as stellar mass is fixed, 
the galaxies are divided into
low-metallicity and high-metallicity groups
using a divider of $12+\mathrm{log(O/H)}=8.5$.
The value corresponds to $\sim0.5\,Z_\odot$
if solar oxygen abundance is assumed to be 
$12+\mathrm{log(O/H)}=8.73$
\citep{2009ARA&A..47..481A}.
Figure~\ref{fig:qpah}a shows that 
the mean $q_\mathrm{PAH}$ value 
of the low-metallicity group (2.3\%) 
is lower than that of the high-metallicity group (3.7\%).
This is in line with the result presented in 
\citet{2014A&A...565A.128C},
that metallicity and PAH fraction are directly linked
in local gas-rich galaxies.
In Figure~\ref{fig:qpah}b, 
$\langle q_\mathrm{PAH} \rangle$ values 
for the different metallicity group of galaxies 
in each stellar mass bin 
(with step size $\Delta\mathrm{log} (M_\mathrm{star}/\mathrm{M}_\odot)=0.5$)
are shown as different symbols.
In the stellar mass range of 
$\mathrm{log} (M_\mathrm{star}/\mathrm{M}_\odot)=[9, 11]$,
$\langle q_\mathrm{PAH} \rangle$ values of 
galaxies in the higher metallicity group 
are always higher than that of galaxies in the lower metallicity group.
Therefore,
it can be concluded that the PAH features are 
correlated with metallicity, 
even if the stellar mass is fixed,
when only two variables (metallicity and stellar mass)
are considered. 

\begin{figure*}
\epsscale{1.15}
\plotone{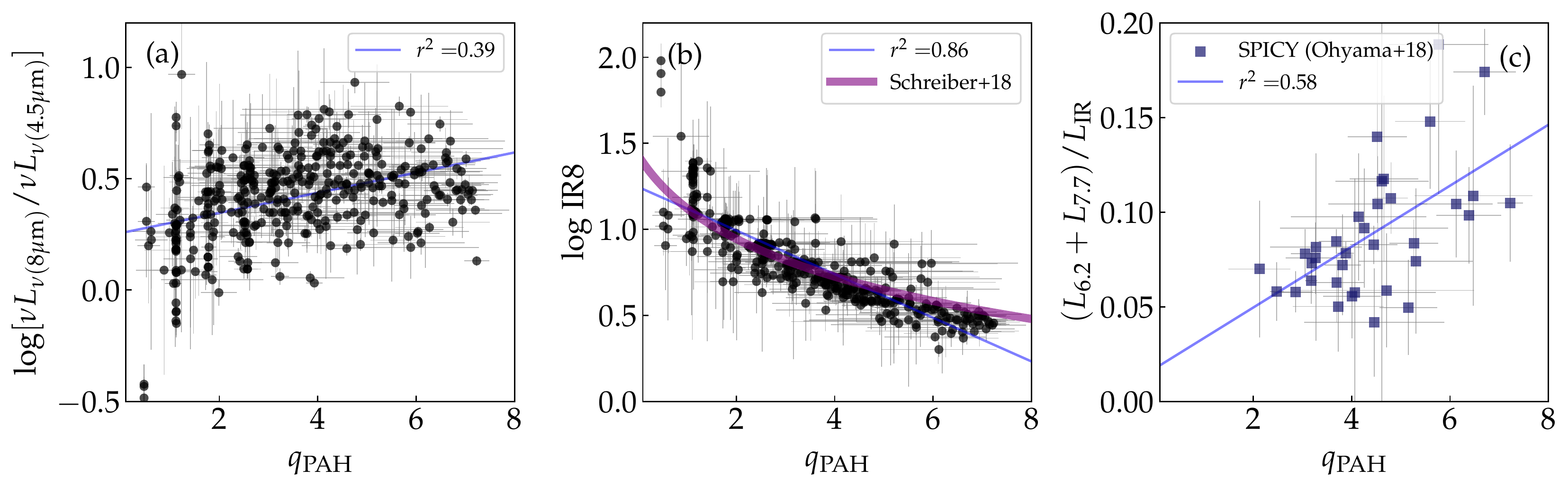}
\caption{\label{fig:qpah_tracer}
	(a) Comparison between 
	$q_\mathrm{PAH}$ derived from SED fitting 
	and 
	luminosity ($\nu L_\nu$) ratio between
        8\,$\mu$m and 4.5\,$\mu$m -- the former
        stands for PAH emission and the latter
        for stellar continuum
	\citep{2014A&A...566A.136M}. 
	Blue solid line is the best-fit linear correlation
	with specified determination coefficient $r^2$. 
	(b) Comparison between 
	$q_\mathrm{PAH}$ and IR8, defined
	as $L_\mathrm{IR}^\mathrm{total}/\nu L_\nu (8\mu\mathrm{m})$.
	Overplotted curve is the 
	correlation between the two suggested by 
	\citet{2018A&A...609A..30S}.
	(c) Comparison between 
	luminosity fraction of PAH features 
	(relative to total IR emission)
	and mass fraction $q_\mathrm{PAH}$. 
	PAH luminosity is calculated by the sum of 
	luminosities of 6.2\,$\mu$m and 7.7\,$\mu$m features
	from \citet{2018A&A...618A.101O}. 
        }
\end{figure*}

From Figure~\ref{fig:figall}, 
it is clear that sSFR 
is strongly correlated 
with the $q_\mathrm{PAH}$, 
showing an even larger correlation coefficient 
than that in case of stellar mass and metallicity. 
The correlation also is in line with 
stellar mass-SFR relation composed of galaxies in 
the star-forming main sequence. 
With symbol colors coded according to sSFR 
in Figures~\ref{fig:qpah}a and \ref{fig:qpah}b, 
the three parameters (metallicity, stellar mass, 
and sSFR) seem to be correlated with each other. 
In order to understand which is
the \textit{primary} correlation 
that drives $q_\mathrm{PAH}$ variation, 
we calculate the partial correlation coefficients
between $q_\mathrm{PAH}$ and each parameter
while the effects of the other two parameters are removed. 
The results are
$0.10$, $0.32$, and $-0.43$ in case of 
$12+\mathrm{log(O/H)}$, $\mathrm{log} M_\mathrm{star}$,
and $\mathrm{log}\,\mathrm{sSFR}$. 
This corresponds to $p$-values of 
0.067, $<10^{-5}$, and $<10^{-5}$. 
All three parameters are correlated to $q_\mathrm{PAH}$
even when effects of other parameters are removed, 
and among these three, 
sSFR appears to be the most strongly correlated 
parameter. 
However, considering the fact that 
coefficient calculation is 
affected by uncertainties as is mentioned above, 
it is still difficult to make a firm conclusion. 
The existence and the effectiveness of 
metallicity-mass-SFR correlation
or so-called fundamental plane of star-forming galaxies
is still being debated 
\citep[e.g.,][]{2010A&A...521L..53L, 2017MNRAS.469.2121S}. 
At least for our sample of MIR-selected galaxies, 
metallicity-mass-SFR correlation does exist
and affects PAH strength.

\subsection{$q_\mathrm{PAH}$ Estimators}  \label{sec:result3}

Using the MIR-selected star-forming galaxies sample,
we compare 
observable luminosity ratios 
suggested by previous studies 
to $q_\mathrm{PAH}$ parameters derived from SED fitting.  
For example, 
\citet{2014A&A...566A.136M} have suggested
8\,$\mu$m-to-4.5\,$\mu$m luminosity ratio traces
PAH emission features,
since rest-frame 8\,$\mu$m luminosity is mainly from PAH emission
and 4.5\,$\mu$m luminosity represents stellar continuum
in case of non-AGN galaxies.
They have shown that 
$\nu L_\nu (8\,\mu\mathrm{m})/\nu L_\nu (4.5\,\mu\mathrm{m})$
increases with ``starburstiness,'' i.e., offset from the
star-forming main sequence, yet the ratio saturates
above a certain limit.
We show in Figure~\ref{fig:qpah_tracer}a, 
how $\nu L_\nu (8\,\mu\mathrm{m})/\nu L_\nu (4.5\,\mu\mathrm{m})$ 
differs as a function of $q_\mathrm{PAH}$. 
Little correlation is found
between $q_\mathrm{PAH}$ and
$\nu L_\nu (8\mu\mathrm{m})/\nu L_\nu (4.5\mu\mathrm{m})$,
suggesting that although
the ratio between 8-$\mu$m and 4.5-$\mu$m luminosity
may give insight to the existence of PAH emission features,
it is difficult to estimate
the relative strength of PAH emission
compared to the total dust emission.

On the other hand,
the IR8 value
\citep[$\mathrm{IR8}\equiv L_\mathrm{IR}/\nu L_\nu (8\mu\mathrm{m})$, ][]{2011A&A...533A.119E}
is well correlated with the $q_\mathrm{PAH}$
(Figure~\ref{fig:qpah_tracer}b). 
The correlation between 
IR8 and $q_\mathrm{PAH}$ 
used in generating a wide range of IR SEDs in
\citet{2018A&A...609A..30S}
is consistent with our sample of galaxies. 

Finally, we show the relative luminosity ratio 
between PAH and total IR luminosity 
versus PAH mass fraction 
in Figure~\ref{fig:qpah_tracer}c, 
using sources for which 5-13\,$\mu$m spectra 
have been obtained \citep{2018A&A...618A.101O}. 
The ratio between
PAH luminosity estimated from specific features 
(rest frame 6.2 and 7.7\,$\mu$m) 
and total IR luminosity 
is positively correlated with $q_\mathrm{PAH}$.
However, scatters imply that 
variation between the PAH features
at different wavelengths 
might be nonnegligible 
in estimating $q_\mathrm{PAH}$. 

\begin{figure*}
\epsscale{1.1}
\plottwo{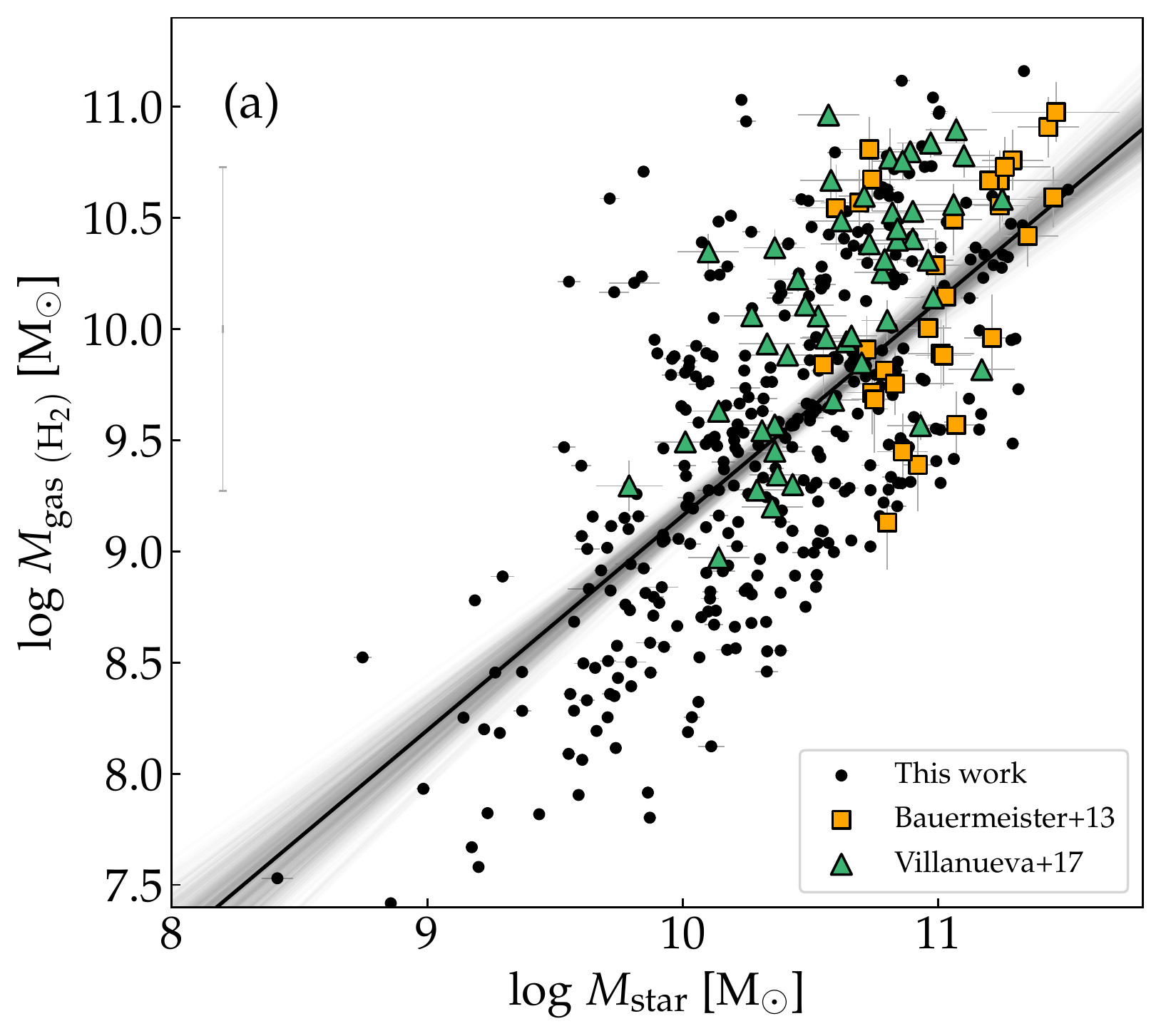}{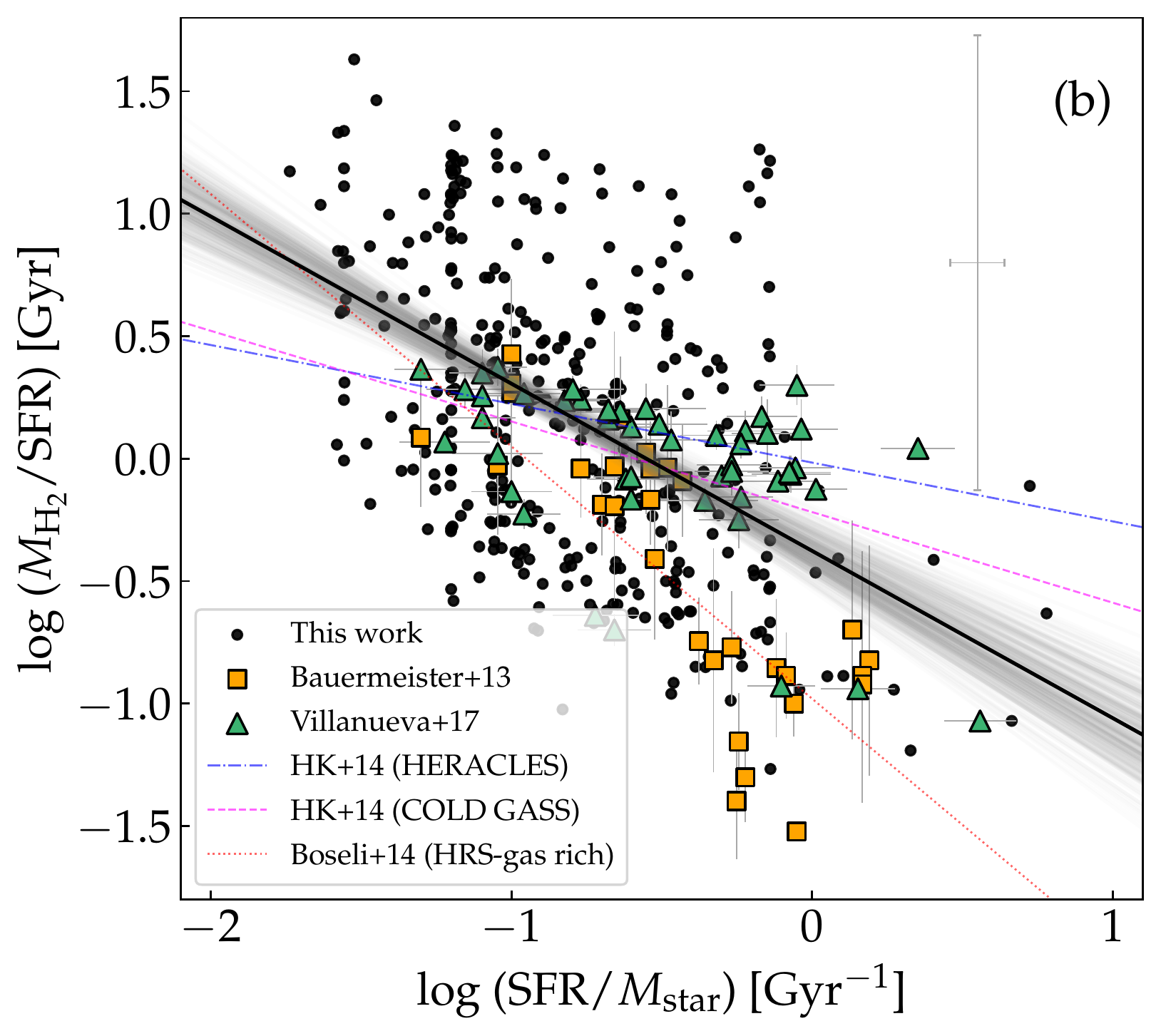}
\caption{\label{fig:mgas}
        (a) Molecular gas mass for MIR-selected galaxies
        as a function of stellar mass.
        In order to calculate the molecular gas mass
        ($M_{\mathrm{H}_2}$),
        total gas mass is estimated using
        $\delta_\mathrm{GDR}$-metallicity calibration
        \citep{2019A&A...623A...5D},
        dust mass from SED fitting,
        and metallicity-dependent
        H$_2$/H\textsc{i} mass ratio
        \citep{2014A&A...564A..66B}.
        Error bar in the left corner represents typical
        uncertainties in the derived $M_{\mathrm{H}_2}$.
        Overplotted symbols are
        $0.05<z<0.5$ star-forming galaxies
        with measured molecular gas mass through CO observations
        \citep[squares and triangles;][]{2013ApJ...768..132B, 2017MNRAS.470.3775V}.
        Linear regression fit between the gas mass and stellar mass
        of our sample galaxies
        is plotted as a thick black line,
        along with thin gray lines
        covering a possible range of linear relation.
        (b) Gas depletion time ($M_\mathrm{gas}/\mathrm{SFR}$)
        vs. sSFR ($\mathrm{SFR}/M_\mathrm{star}$)
        for our sample of MIR-selected galaxies.
        Error bar in the corner represents
        typical uncertainties in $x$- and $y$-axis parameters,
        reflecting the estimation errors as well as
        scatter in a conversion formula.
        Again, the CO-detected samples from
        \citet{2013ApJ...768..132B} and \citet{2017MNRAS.470.3775V}
        are plotted for comparison.
        The thick black line and thin gray lines show
        linear regression results.
        Dot-dashed and dashed lines indicate
        depletion time vs. sSFR relation
        for nearby ($z<0.05$) galaxies
        measured either using global gas mass
        \citep[COLD GASS;][]{2011MNRAS.415...61S, 2014MNRAS.443.1329H}
        or gas mass derived for 1\,kpc grids
        \citep[HERACLES;][]{2014MNRAS.443.1329H}.
        Dotted line is such a relation
        for local (with distance $<25$\,Mpc) `gas-rich' galaxies
        among the Herschel reference survey galaxies
        \citep[][]{2014A&A...564A..66B}.
        }
\end{figure*}

\subsection{Dust Mass as a Tracer of Gas Mass}  \label{sec:result4}

It is suggested that total dust mass ($M_\mathrm{dust}$) 
in a galaxy 
can be used as a proxy
for estimating gas mass  
\citep[e.g.,][]{2015ApJ...799...96G, 2016A&A...587A..73B},
since the gas and dust content of galaxies 
are linked to each other 
through a gas-to-dust mass ratio ($\delta_\mathrm{GDR}$). 
The gas-to-dust ratio is found to be 
a function of metallicity, in a way that 
gas-to-dust ratio decreases as the metallicity increases
\citep[][]{2011ApJ...737...12L, 2012ApJ...760....6M, 
2014A&A...563A..31R, 2019A&A...623A...5D}.
For our MIR-selected star-forming galaxies at $z<0.4$,
we combine the metallicity-dependent gas-to-dust ratio
($\delta_\mathrm{GDR}$)
with the dust mass ($M_\mathrm{dust}$) 
to estimate the gas mass in these galaxies
($M_\mathrm{gas}\equiv\delta_\mathrm{GDR}M_\mathrm{dust}$).
Dust masses are estimated from the SED fitting 
(Section~\ref{sec:sedfit}) using the dust emission model of
\citet{2007ApJ...663..866D}.
For gas-to-dust ratio, we use the following 
$\delta_\mathrm{GDR}$-metallicity calibration
from \citet{2019A&A...623A...5D}: 

\begin{equation}
        \mathrm{log}\delta_\mathrm{GDR}=(21.19\pm0.90)-(2.15\pm0.11)(12+\mathrm{log(O/H)}).
\end{equation}

The formula is calibrated using Dustpedia data sets
that consist of nearby ($v<3000$\,km s$^{-1}$) 
extended galaxies with Herschel FIR observations. 
Differences in the dust emission models 
used in \citet{2019A&A...623A...5D} 
\citep[THEMIS;][]{2019A&A...624A..80N} 
and our work \citep{2007ApJ...663..866D} are taken 
into account 
(for systematics and uncertainties of measuring dust mass
based on different methods of modeling FIR dust emission,
see \citealt{2016A&A...587A..73B}). 
In our case, using the \citet{2007ApJ...663..866D} model
leads to a factor of 1.5 higher dust mass
compared to the case of using THEMIS model.
The gas mass in the calibration formula 
corresponds to the `total' gas mass, including 
molecular hydrogen and atomic hydrogen,
as well as gas of heavier elements 
\citep[$M_\mathrm{gas}=\xi M_\mathrm{H\textsc{i}}(1+M_\mathrm{H_2}/M_\mathrm{H\textsc{i}})$,][]{2019A&A...623A...5D}.
Therefore, in order to compare our results
with CO observations of star-forming galaxies
at similar redshift range, 
we calculate molecular hydrogen mass 
by applying $\xi=1.39$ and 
$M_\mathrm{H_2}/M_\mathrm{H\textsc{i}}$ ratio
that is dependent on the oxygen abundance  
\citep[][]{2014A&A...564A..66B}.
In our oxygen abundance range, 
the $M_\mathrm{H_2}/M_\mathrm{H\textsc{i}}$ varies
in the range of 0.1 to 0.8.

Figure~\ref{fig:mgas}a shows that
our MIR-selected galaxies have molecular hydrogen gas mass
of the range of $\mathrm{log}\,M_\mathrm{H_2} [\mathrm{M}_\odot]=7.5$-11.5. 
Spearman rank coefficient between $\mathrm{log}\,M_\mathrm{H2}$
and $\mathrm{log}\,M_\mathrm{star}$ is $0.60$,
and if the errors in two parameters are taken into account, 
the Pearson correlation coefficient is $0.91\pm0.04$.
The linear regression suggests following correlation
between the stellar and dust mass: 

\begin{equation}
	\mathrm{log}\,M_\mathrm{H_2} [\mathrm{M}_\odot] = (-0.49\pm0.63)+(0.96\pm0.06)~\mathrm{log}\,M_\mathrm{star} [\mathrm{M}_\odot].
\end{equation}

Compared in Figure~\ref{fig:mgas}a are star-forming galaxies
at similar redshift range
\citep[$0.05<z<0.5$;][]{2013ApJ...768..132B, 2017MNRAS.470.3775V},
of which gas masses are estimated from CO observations.
Galaxies from \citet{2013ApJ...768..132B}
(squares in Figure~\ref{fig:mgas}a)
are selected based on the large SFR
(of order of several tens $\mathrm{M}_\odot\,\mathrm{yr}^{-1}$)
with a stellar mass of order of $10^{10.6-11.5}\,\mathrm{M}_\odot$,
thus are galaxies above the star-forming galaxy main sequence
in SFR-stellar mass plane,
sharing similar positions with local LIRGs.
Galaxies from \citet{2017MNRAS.470.3775V}
(triangles in Figure~\ref{fig:mgas}a)
are galaxies selected in the 160\,$\mu$m,
of which total IR luminosity is larger than
$10^{11}\,\mathrm{L}_\odot$ in most cases.
Therefore the compared $0.05<z<0.5$ galaxies with CO detections
are `scaled-up' versions of our MIR-selected galaxies
with an order of larger average stellar mass and an order of larger SFR.
The fact that CO-detected galaxies 
show similar $M_\mathrm{H_2}$-$M_\mathrm{star}$ correlation with 
our sample 
implies that the gas-to-star mass fraction 
is roughly consistent 
in a wide stellar mass range of $10^9$-$10^{11}$\,M$_\odot$.

In Figure~\ref{fig:mgas}b,
we show the molecular gas depletion time
($\tau_\mathrm{dep}\equiv M_\mathrm{H_2}/\mathrm{SFR}$)
of our sample galaxies
as a function of sSFR.
From the CO-detected galaxies 
\citep{2013ApJ...768..132B, 2017MNRAS.470.3775V},
it is noticeable that the gas depletion time scale
has a negative correlation with the sSFR. 
Such a trend is also 
observed for nearby ($z<0.05$) galaxies
\citep[see dashed, dotted-dashed, and dotted lines
in Figure~\ref{fig:mgas}b; ][]{2014MNRAS.443.1329H, 2014A&A...564A..66B},
and implies that a fast consumption of gas in starburst galaxies
is present throughout to $z\sim0.5$.
Although the uncertainties in 
gas mass (and thus $\tau_\mathrm{dep}$) is very large, 
we derive the following relation between $\tau_\mathrm{dep}$ 
and sSFR through linear regression: 

\begin{equation}
	\mathrm{log\,\tau_\mathrm{dep}} [\mathrm{Gyr}] = (-0.38\pm0.06)+(-0.68\pm0.07)~\mathrm{log\,sSFR} [\mathrm{Gyr}^{-1}].
\end{equation}

Note that Spearman rank coefficient between two parameters 
is $-0.45$, while the linear correlation coefficient
estimated with the consideration of the parameter uncertainties
is $-0.88\pm0.07$.
Like local and low-redshift galaxies with CO detection, 
MIR-selected galaxies are suspected to have 
shorter gas depletion time scales 
if the sSFR are larger. 
\citet{2014A&A...564A..66B} have suggested that
the slope of the $\tau_\mathrm{dep}$-sSFR relation
is steeper if only gas-rich, late type galaxies are considered, 
compared to the case when all galaxies
(targets of CO observation with only upper limits, 
i.e., nondetections, 
like in the case of \citet{2014MNRAS.443.1329H})
are used to derive the correlation.
The $\tau_\mathrm{dep}$-sSFR relation of 
our MIR-selected sample galaxies 
suggests that MIR-selected galaxies are 
similar to gas-rich galaxies.

\section{Summary}  \label{sec:summary}

Based on the optical spectra 
and the multiwavelength photometry, 
we estimate gas-phase metallicity 
and physical parameters of MIR-selected galaxies at $z<0.4$. 
The motivation of this work 
is to utilize multiple broad-band photometric information 
in MIR wavelengths 
to constrain relative strength of PAH emission, 
which is a key to understanding the scatter 
in ratio between the MIR and total IR luminosity.  
The total IR luminosity as well as the mass fraction of 
PAH particles among total dust mass 
are estimated through SED fitting. 

The PAH mass fraction $q_\mathrm{PAH}$
is dependent on the gas-phase metallicity 
measured using strong emission lines, 
i.e., $q_\mathrm{PAH}$ increases as the metallicity increases. 
The result is in line with previous studies 
using local star-forming galaxies. 
Since possible AGN contaminants are excluded from the sample
based on optical line ratios and MIR colors, 
the result shows that characteristics of low metallicity 
ISM environments 
for pure H\,\textsc{ii} galaxies 
suppress the PAH emission. 
Suggestions have been made that 
a harder UV-radiation field (originating from less dust shielding) 
in a low-metallicity environment 
is effective in the destruction of PAH particles. 
The $q_\mathrm{PAH}$ value is also dependent 
on the stellar mass and sSFR,
in a way that PAH features are weaker 
in less massive galaxies
and in galaxies where current SFR is 
relatively larger compared to the accumulated stellar mass. 
Partial correlation coefficients
between $q_\mathrm{PAH}$ and physical parameters
suggest that 
$q_\mathrm{PAH}$ value of galaxies
with similar stellar mass and sSFR 
is still a function of metallicity, 
while the same trend is found for the cases of
stellar mass and sSFR. 
This reflects the mass-metallicity relation of galaxies, 
as well as implying the mass-metallicity-SFR scaling relation
of star-forming galaxies. 

The metallicity of galaxies not only affects 
the PAH strength that dominates MIR wavelengths,
but also provides a link between gas and dust mass 
in galaxies. 
Applying metallicity-dependent gas-to-dust ratio calibration 
to our sample galaxies, 
a trend of decreasing gas depletion time scale
along the increase of sSFR 
is reproduced,
similarly to the case of CO-observed galaxies. 
Considering these, metallicity is one of the main factors
that is necessary to describe
the variation in MIR-FIR SED.

\begin{acknowledgments}

H.S. acknowledges support from the National Research Foundation of Korea (NRF)
grant Nos. 2021R1A2C4002725 and 2022R1A4A3031306, 
funded by the Korea government (MSIT).
H.S.H. acknowledges the support by the National Research Foundation of Korea (NRF) 
grant funded by the Korea government (MSIT; No. 2021R1A2C1094577). 
D.K. acknowledges support by the National Research Foundation of Korea (NRF) grant 
(No. 2021R1C1C1013580) funded by the Korean government (MSIT). 
T.H. acknowledges the support of the National Science and Technology Council of Taiwan
through grants 110-2112-M-005-013-MY3, 110-2112-M-007-034-, and 111-2123-M-001-008-.
T.N. acknowledges the support by JSPS KAKENHI grant No.
21H04496.
A.N. acknowledges support from the Narodowe Centrum Nauki (UMO-2020/38/E/ST9/00077).
W.J.P. has been supported by the Polish National Science Center (NCN) project 
UMO-2020/37/B/ST9/00466 and the Foundation for Polish Science (FNP).
This research is based on observations with AKARI, a JAXA project with
the participation of ESA.
Some of the data presented in this paper were obtained    
from the Mikulski Archive for Space Telescope (MAST)  
at the Space Telescope Science Institute. 
The specific observations analyzed can be accessed via
\dataset[10.17909/T9H59D]{https://doi.org/10.17909/T9H59D}.

\end{acknowledgments}

%

\vspace{5mm}
\facilities{AKARI (IRC), Herschel (SPIRE), MMT (Hectospec), WIYN (Hydra), 
GALEX, WISE, Spitzer (IRAC, MIPS), Subaru (HSC), CFHT (Megacam)}


\software{CIGALE \citep{2019A&A...622A.103B},
	  MPFIT \citep{2009ASPC..411..251M},
	  XID+ \citep{2017MNRAS.464..885H}
	  }

\end{document}